\documentclass[journal]{IEEEtran}
\usepackage[cmex10]{amsmath}
\usepackage{graphicx,amssymb,amsthm,comment,bm,cite,url}
\usepackage{color}
\usepackage{boldline}
\usepackage{algorithm,algorithmic}
\usepackage{tabularx,adjustbox}
\usepackage{multirow}
\allowdisplaybreaks

\newcommand{\refeqs}[2]{Eqs. (\ref{eq:#1}) and (\ref{eq:#2})}

\newcommand{\refsubsec}[1]{Subsection \ref{subsec:#1}}

\newcommand{\reffig}[1]{Fig. \ref{fig:#1}}
\newcommand{\reffigs}[2]{Figs. \ref{fig:#1} and \ref{fig:#2}}

\newcommand{\reftab}[1]{Table \ref{tab:#1}}

\def\Vec#1{\boldsymbol{\mathbf{#1}}}

\def\thline{\noalign{\hrule height 1.2pt}}

\ifCLASSINFOpdf
\else
\fi

\begin{document}
\title{
FastS2S-VC: 
Streaming
Non-Autoregressive 
Sequence-to-Sequence Voice Conversion
}

\author{Hirokazu~Kameoka,
Kou Tanaka, and
Takuhiro Kaneko%
\thanks{H. Kameoka, K. Tanaka, and T. Kaneko
are with NTT Communication Science Laboratories, Nippon Telegraph and Telephone Corporation, Atsugi, Kanagawa, 243-0198 Japan (e-mail: hirokazu.kameoka.uh@hco.ntt.co.jp).}% <-this % stops a space
}

\markboth{}%
{}

\maketitle

\begin{abstract}
This paper proposes a non-autoregressive extension of our previously proposed sequence-to-sequence (S2S) model-based voice conversion (VC) methods. 
S2S model-based VC methods have attracted particular attention in recent years for their flexibility in converting not only the voice identity but also the pitch contour and local duration of input speech, thanks to the ability of the encoder-decoder architecture with the attention mechanism. 
However, one of the obstacles to making these methods work in real-time is the autoregressive (AR) structure.
To overcome this obstacle, we develop a method to obtain a model that is free from an AR structure and behaves similarly to the original S2S models, based on a teacher-student learning framework.
In our method, called ``FastS2S-VC'', 
the student model consists of encoder, decoder, and attention predictor. 
The attention predictor learns to predict attention distributions solely from source speech along with a target class index with the guidance of those predicted by the teacher model from both source and target speech. 
Specifically, it is designed as a network that takes the mel-spectrogram of source speech as input and generates in parallel each row of an attention weight matrix represented by a Gaussian function.
Thanks to the model structure that does not take the mel-spectrogram of target speech as input, the model is freed from an AR structure and allows for parallelization.
%These Gaussians are constrained to be placed in left-to-right order so as to ensure that the attended time point always move forward. 
Furthermore, we show that FastS2S-VC is suitable for real-time implementation based on a sliding-window approach, and describe how to make it run in real-time.
Through speaker-identity and emotional-expression conversion experiments, we confirmed that FastS2S-VC was able to speed up the conversion process by 70 to 100 times compared to the original AR-type S2S-VC methods, without significantly degrading the audio quality and similarity to target speech.
We also confirmed that the real-time version of FastS2S-VC can be run with a latency of 32 ms when run on a GPU.
\end{abstract}


\IEEEpeerreviewmaketitle

\section{Introduction}
\label{sec:intro}

The technique to 
modify some characteristics of speech, such as voice identity, emotional expression, and accents,
without changing the linguistic content is called voice conversion (VC).
Early studies of VC focused on learning spectral feature mapping rules using time-aligned parallel utterances of source and target speech. 
To name just a few, examples include methods based on Gaussian mixture models (GMMs) \cite{Stylianou1998short,Toda2007short}, partial least square regression \cite{Helander2010}, frequency warping \cite{Tian2017}, non-negative matrix factorization \cite{Takashima2012}, group sparse representation \cite{Sisman2017}, fully connected deep neural networks (DNNs) \cite{Desai2010short,Mohammadi2014short}, and long-short term memory networks (LSTMs) \cite{Sun2015,Ming2016lstm}.
Along with these studies, there has also been a lot of work done on non-parallel VC methods, which require no parallel utterances for training.
These include methods based on an i-vector representation \cite{Kinnunen2017}, restricted Boltzmann machines \cite{Nakashika2014}, phonetic posteriograms \cite{Sun2016}, 
regular/variational autoencoders \cite{Hsu2016short, Hsu2017short, vandenOord2017, YSaito2018b, Kameoka2019IEEETransshort_ACVAE-VC, Tobing2019, Qian2019}, generative adversarial networks \cite{Kaneko2018, Kameoka2018SLT_StarGAN-VC}, flow-based models \cite{Serra2019}, and score-based generative models \cite{Kameoka2020_VoiceGrad}. 
More details on the trends and challenges in 
VC research can be found in a recent review article \cite{Sisman2020}.

Among many VC studies, methods based on the sequence-to-sequence (S2S) learning framework \cite{Sutskever2014short,Chorowski2015NIPS} have received particular attention in recent years \cite{Miyoshi2017short,Zhang2018short,Zhang2019short,Biadsy2019short,Liu2020,Tanaka2019ICASSP_AttS2S-VC,Kameoka2018arXiv_ConvS2S-VC,Kameoka2018IEEE-TASLP_ConvS2S-VC,Huang2020Interspeech_VTN,Kameoka2020IEEE-TASLP_VTN}. The S2S learning framework offers a general and powerful solution to a broad class of sequential mapping problems, including machine translation (MT), automatic speech recognition (ASR) \cite{Chorowski2015NIPS}, and text-to-speech (TTS) \cite{Wang2017short,Arik2017ashort,Arik2017bshort,Sotelo2017short,Tachibana2018short,Ping2018ICLRshort}.
While most conventional VC methods have focused on converting only the voice identity, S2S-based VC (hereafter referred to as S2S-VC) methods are particularly attractive in that they can flexibly change not only the voice identity but also the speaking style, thanks to the ability of the encoder-decoder architecture with the attention mechanisim.
However, one challenge in S2S-VC is how to make the model work well even with a limited amount of training data given the high cost of collecting parallel utterances.
One idea involves using auxiliary text transcriptions for training, as in \cite{Miyoshi2017short,Zhang2018short,Zhang2019short,Biadsy2019short}. 
As another way, we proposed methods with recurrent neural network (RNN) \cite{Tanaka2019ICASSP_AttS2S-VC}, convolutional neural network (CNN) \cite{Kameoka2018arXiv_ConvS2S-VC,Kameoka2018IEEE-TASLP_ConvS2S-VC}, and Transformer \cite{Huang2020Interspeech_VTN,Kameoka2020IEEE-TASLP_VTN} architectures,
and several ideas to allow these architectures to work stably under limited resources.

When building real-time S2S-VC systems, another important challenge is how to keep the latency of the conversion process as low as possible.
One obstacle to this is the autoregressive (AR) structure common to S2S models.
Hence, one solution would be to develop a non-autoregressive (NAR) S2S model tailored to VC.
Recently, several attempts have already been made to develop NAR-S2S models designed for MT \cite{Gu2018}, ASR \cite{NChen2020}, and TTS \cite{Ren2019NIPS,Ren2020,Elias2020}.
Inspired by these studies, in this paper we propose ``FastS2S-VC'', a VC method based on a VC-tailored NAR-S2S model built upon our convolutional S2S (ConvS2S)-based \cite{Kameoka2018arXiv_ConvS2S-VC,Kameoka2018IEEE-TASLP_ConvS2S-VC} and Transformer-based \cite{Huang2020Interspeech_VTN,Kameoka2020IEEE-TASLP_VTN} models, and describe its real-time implementation. 

\section{Sequence-to-Sequence Voice Conversion}
\label{sec:S2S-VC}

\subsection{General Structure}
\label{subsec:GeneralForm}

In this section, we first introduce a general form of our S2S-VC models, which reduces to the architectures we proposed previously \cite{Tanaka2019ICASSP_AttS2S-VC,Kameoka2018arXiv_ConvS2S-VC,Kameoka2018IEEE-TASLP_ConvS2S-VC,Huang2020Interspeech_VTN,Kameoka2020IEEE-TASLP_VTN} under certain designs.

We hereafter use
$\Vec{X}^{(k)}=[\Vec{x}_1^{(k)},\ldots,
\Vec{x}_{N}^{(k)}]\in \mathbb{R}^{D\times N}$ and 
$\Vec{X}^{(k')}=[\Vec{x}_1^{(k')},\ldots,
\Vec{x}_{M}^{(k')}]\in \mathbb{R}^{D\times M}$
to denote the feature vector sequences 
of source and target speech reading the same sentence,
where the source and target speech are assumed to belong to classes $k$ and $k'$. 
$N$ and $M$ denote the lengths of the two sequences and $D$ denotes the feature dimension. 
Here, a class represents any non-linguistic attribute of speech, for example, speaker identity in a speaker conversion task and emotional state in an emotional expression conversion task. 
In the following, we consider a many-to-many S2S model that learns to map $\Vec{X}^{(k)}$ to $\Vec{X}^{(k')}$ where $k,k'\in\{1,\ldots,K\}$.

While in our previous studies, we chose to use a set of the mel-cepstral vocoder parameters (the mel-cepstral coefficients, log fundamental frequency, and aperiodicity) extracted at each short-term frame as the feature vector  \cite{Tanaka2019ICASSP_AttS2S-VC,Kameoka2018arXiv_ConvS2S-VC,Kameoka2018IEEE-TASLP_ConvS2S-VC,Kameoka2020IEEE-TASLP_VTN}, this was only a tentative choice 
%before using one of neural vocoders for waveform generation. 
in anticipation of the future use of one of neural vocoders for waveform generation.  
In this paper, we use the 80-dimensional mel-spectrum instead as the feature vector and choose to use Parallel WaveGAN \cite{Yamamoto2020} for waveform generation.
In the following, for the sake of distinction, we refer to these versions of RNNS2S-VC \cite{Tanaka2019ICASSP_AttS2S-VC}, ConvS2S-VC \cite{Kameoka2018arXiv_ConvS2S-VC,Kameoka2018IEEE-TASLP_ConvS2S-VC}, and Transformer-VC (also called Voice Transformer Network; VTN) \cite{Huang2020Interspeech_VTN,Kameoka2020IEEE-TASLP_VTN} as RNNS2S-VC2, ConvS2S-VC2, and Transformer-VC2, respectively.
For the purpose of speeding up and stabilizing the training and inference of S2S-VC models, we split the mel-spectral sequence (mel-spectrogram) of each utterance into non-overlapping segments of equal length $r$ and use the stack of the mel-spectra within each segment as a new feature vector, as in \cite{Kameoka2018IEEE-TASLP_ConvS2S-VC,Huang2020Interspeech_VTN,Kameoka2020IEEE-TASLP_VTN}. 
This reshaping makes the new feature sequence $r$ times shorter than the original mel-spectrogram.
Thus, $D=80\times r$.

\begin{figure}[t!]
\centering
\begin{minipage}[t]{.98\linewidth}
  \centerline{\includegraphics[width=.98\linewidth]{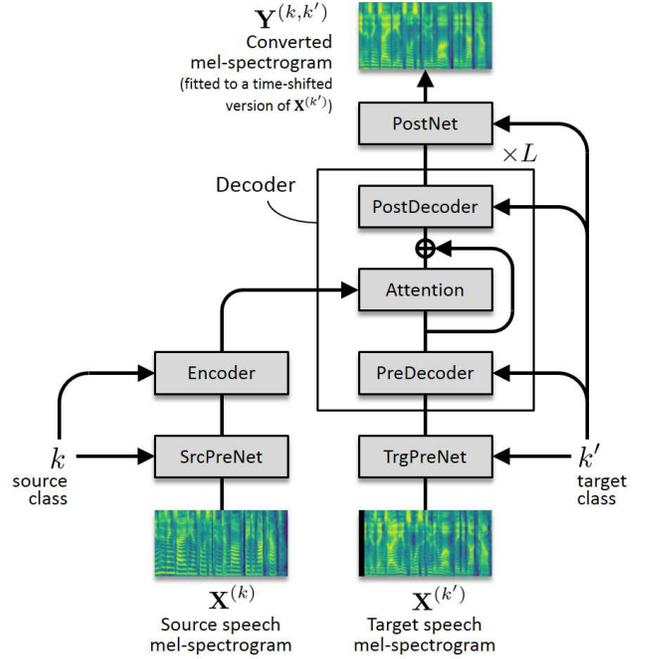}}
  %\vspace{-1ex}
  \caption{General structure of our S2S-VC model for the case of $r=1$. Here, $\oplus$ is used to denote either a sum of arrays with compatible sizes or a concatenation of arrays along the channel axis. The notation ``$\times L$'' indicates $L$ repetitions of the block enclosed by the solid frame. 
  }
\label{fig:GeneralForm}
\end{minipage}
%\vspace{-1ex}
\end{figure}

The overall structure of the general form of our S2S-VC model is illustrated in \reffig{GeneralForm}.
As \reffig{GeneralForm} shows, our S2S-VC model consists of seven modules: 
source and target prenets, encoder, predecoder, attention module, postdecoder, and postnet.
The roles of these modules are as follows.
The two prenets are responsible for capturing the local dynamics in $\Vec{X}{}^{(k)}$ and $\Vec{X}{}^{(k')}$, the mel-spectrograms of source and target speech. 
The encoder and predecoder are expected to extract the linguistic contents of source and target speech in the form of context vector sequences $\Vec{Z}{}^{(k)}$, $\Vec{Z}{}^{(k')}$ from the sequences $\tilde{\Vec{X}}{}^{(k)}$, $\tilde{\Vec{X}}{}^{(k')}$ produced by the two prenets. 
The attention module computes a similarity (attention) matrix $\Vec{A}{}^{(k,k')}$ between the source and target context vector sequences, and uses it to warp the source context vector sequence so that the warped time points contextually correspond to the time points of the target context vector sequence.
The postdecoder and postnet are responsible for converting the warped source context vector sequence $\Vec{R}{}^{(k,k')}$ into a time-shifted version of the target mel-spectrogram so that the final output from the postnet at a certain step can be used recursively as an input into the target prenet at the next step at test inference time. 
For this reason, it is important to note that the target prenet, predecoder, postdecoder, and postnet must be designed in such a manner that they do not use future information of the input sequence when producing the output at each time step. 

\begin{figure*}[t!]
\centering
\begin{minipage}[t]{.98\linewidth}
  \centerline{\includegraphics[width=.98\linewidth]{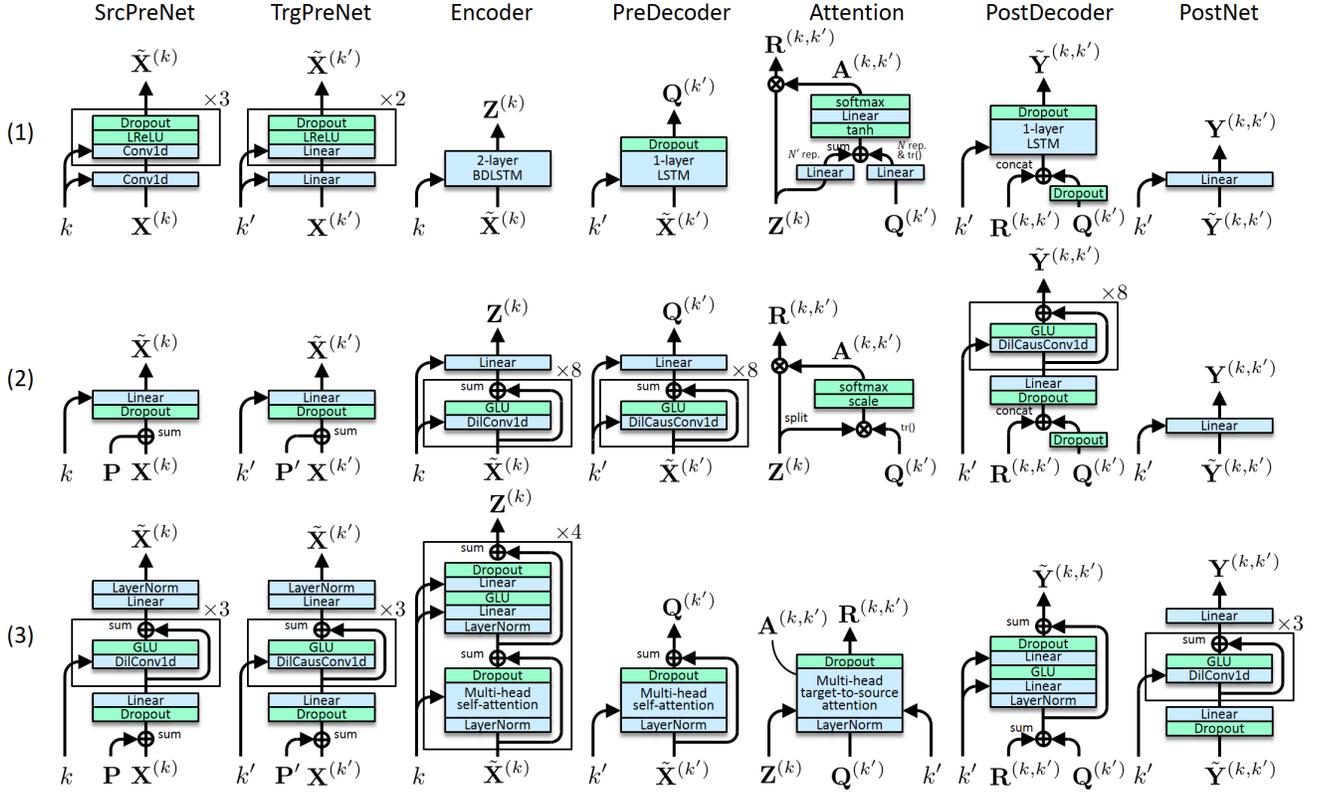}}
  %\vspace{-1ex}
  \caption{Architecture details of (1) RNNS2S-VC2, (2) ConvS2S-VC2, and (3) Transformer-VC2. The blue boxes indicate layers with learnable parameters and the green boxes indicate operations with no learnable parameters. The attention module internally generates one attention weight matrix $\Vec{A}{}^{(k,k')}$ for (1) and (2), and  $H$ attention weight matrices $\Vec{A}{}^{(k,k')}=\{\Vec{A}_h^{(k,k')}\}_{1\le h\le H}$ for (3). In each layer that takes a class index as an input, an embedding vector is first retrieved from a learnable lookup table according to that class index, then the embedding vector is repeated along the time axis, and the resulting vector sequence is finally appended to the input sequence along the channel direction before being fed into that layer. $\Vec{P}$ and $\Vec{P}'$ denote sinusoidal position encodings \cite{Vaswani2017NIPSshort}.
  }
\label{fig:arch}
\end{minipage}
%\vspace{-1ex}
\end{figure*}

Note that in most literatures related to S2S models, the sub-modules that we term 
the predecoder, attention module, and postdecoder 
in the above model are often collectively called the ``decoder''.
However, we will stick to this sub-module representation for the sake of brevity in the subsequent explanations.
In fact, it allows us to view the architectures of RNNS2S-VC2, ConvS2S-VC2, and Transformer-VC2 in a unified manner as special cases of the above model.
\reffig{arch} shows the designs of the seven modules where the above model reduces to the RNNS2S-VC2, ConvS2S-VC2, and Transformer-VC2 models.

\subsubsection{RNNS2S-VC2}
The architecture of RNNS2S-VC \cite{Tanaka2019ICASSP_AttS2S-VC} was inspired by ``Tacotron'' \cite{Wang2017short}, an S2S model-based end-to-end TTS system. The same applies to RNNS2S-VC2, except for some slight modifications. 
The source prenet is composed of four 1D convolution layers with the kernel size of 5, each followed by a leaky rectified linear unit (ReLU) activation function. 
The target prenet is composed of three fully-connected linear layers (applied independently to each vector in an input sequence), each followed by a leaky ReLU activation function.
The encoder is a two-layer birectional LSTM whereas the predecoder is a one-layer unidirectional LSTM. 
The attention module computes Bahdanau's additive attention \cite{Bahdanau2015}. 
The postdecoder is again a one-layer unidirectional LSTM and the postnet is a single fully-connected linear layer that projects the vectors produced by the postdecoder onto mel-spectra. 
Note that the LSTMs in the predecoder and postdecoder are unidirectional so as not to allow them to look at future information in computing the outputs.
Weight normalization \cite{Salimans2016short} is applied to all the learnable weights.
$L$ is 1.
The source and target class labels are fed into each module in the form of continuous vectors stored in a learnable lookup table.

\subsubsection{ConvS2S-VC2}

The architecture of ConvS2S-VC2 is also slightly modified from the original architecture \cite{Kameoka2018arXiv_ConvS2S-VC,Kameoka2018IEEE-TASLP_ConvS2S-VC}.
The source and target prenets and postnet are each composed of a single fully-connected linear layer.
The encoder is composed of eight dilated convolution layers with the kernel size of 5 and the dilation factors of 1, 3, 9, 27, 1, 3, 9, and 27, respectively, each followed by a gated linear unit (GLU) activation function \cite{Dauphin2017short}. 
The same design is used for the predecoder and postdecoder except that all convolutions are causal so as not to allow them to look at future information in computing the outputs.
The attention module computes the scaled dot-product attention introduced in \cite{Vaswani2017NIPSshort}.
$L$ is 1.
Weight normalization \cite{Salimans2016short} is applied to all the learnable weights.

\subsubsection{Transformer-VC2}

The architecture of Transformer-VC2 is exactly the same as the one we proposed in \cite{Kameoka2020IEEE-TASLP_VTN},  except that it uses the mel-spectrograms of source and target speech as the input and output. See \reffig{arch} for the architecture details.

\subsection{Autoregressive Structure}

In all the above models, the target prenet is assumed to take the output vector generated by the postnet at the previous recursive step as the input at each step during test time. This implies that the process of generating the mel-spectrogram of target speech cannot be parallelized. 
This AR structure, which is common in all regular S2S models, is a major bottleneck for achieving real-time VC. Hence, the challenge is how to achieve a model that is free from an AR structure without compromising the capabilities of S2S models.

\subsection{Real-Time System Design}
\label{subsec:S2S-VC_RT}

As already described in \cite{Kameoka2018IEEE-TASLP_ConvS2S-VC,Kameoka2020IEEE-TASLP_VTN}, 
it is actually possible to implement real-time systems with several slight modifications to the above models:
First, we make the source prenet and encoder depend only on past and present inputs as with the target prenet, predecoder, and postdecoder. 
This can be done by replacing all the convolutions and LSTMs with causal and unidirectional versions, respectively, and applying appropriately designed masks to all the self-attention matrices so that position $n$ will depend only on the elements at positions less than $n$.
Second, we design the postdecoder so that it does not directly receive the output from the predecoder (i.e., $\Vec{Q}^{(k')}$ in \reffig{arch}). 
Namely, we let the postdecoder receive only the ouput of the attention module and the target class index (i.e., $\Vec{R}^{(k,k')}$ and $k'$ in \reffig{arch}).
Third, at test time, we force the attention weight matrices to be identity matrices.
This frees the models from the AR structure and makes inference faster and easier. 

\begin{figure*}[t!]
\centering
\begin{minipage}[t]{.49\linewidth}
\centering
\centerline{\includegraphics[width=.98\linewidth]{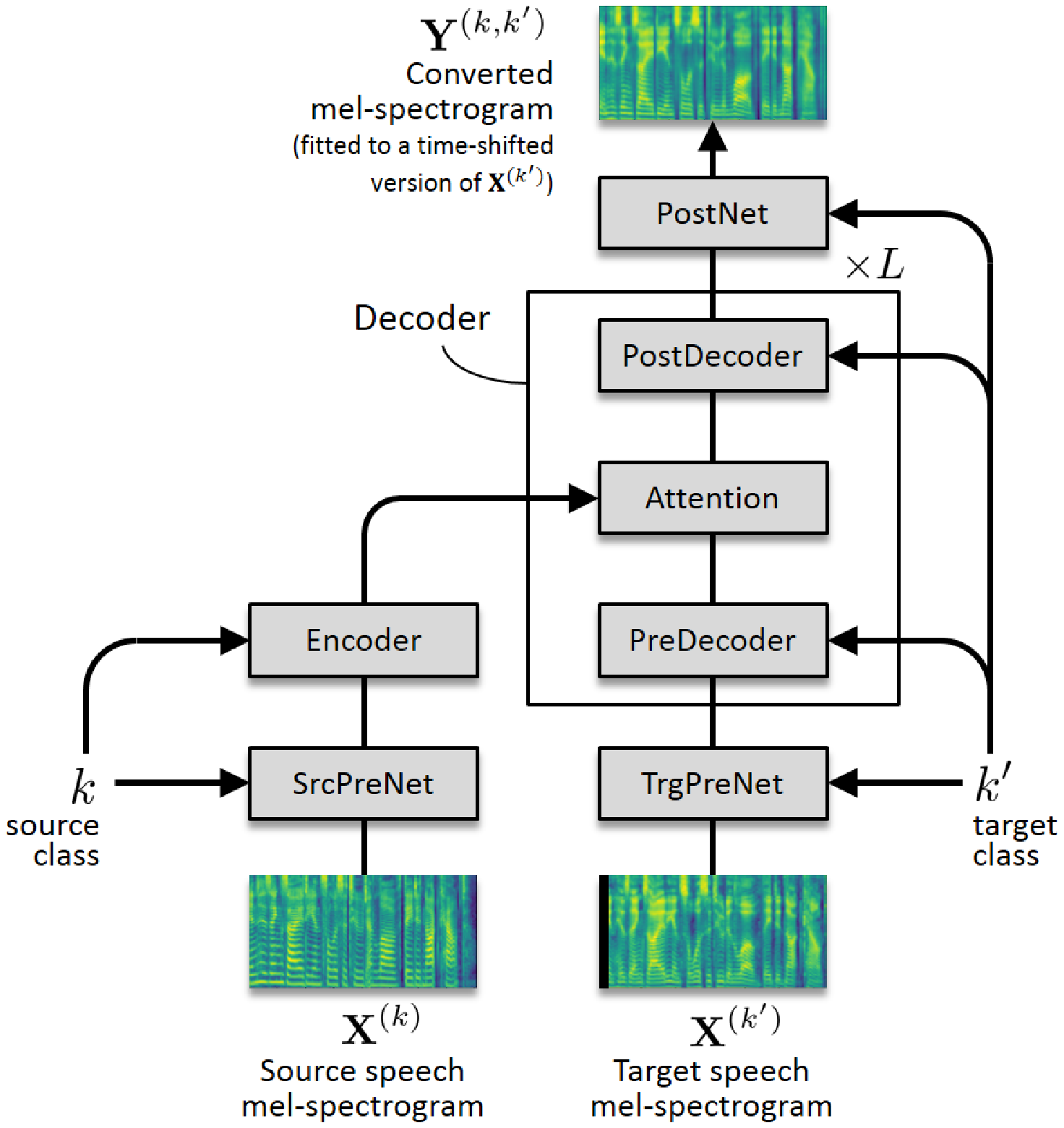}}
%\vspace{-1ex}
{\footnotesize (a) Teacher model}
\end{minipage}
\begin{minipage}[t]{.49\linewidth}
\centering
\centerline{\includegraphics[width=.98\linewidth]{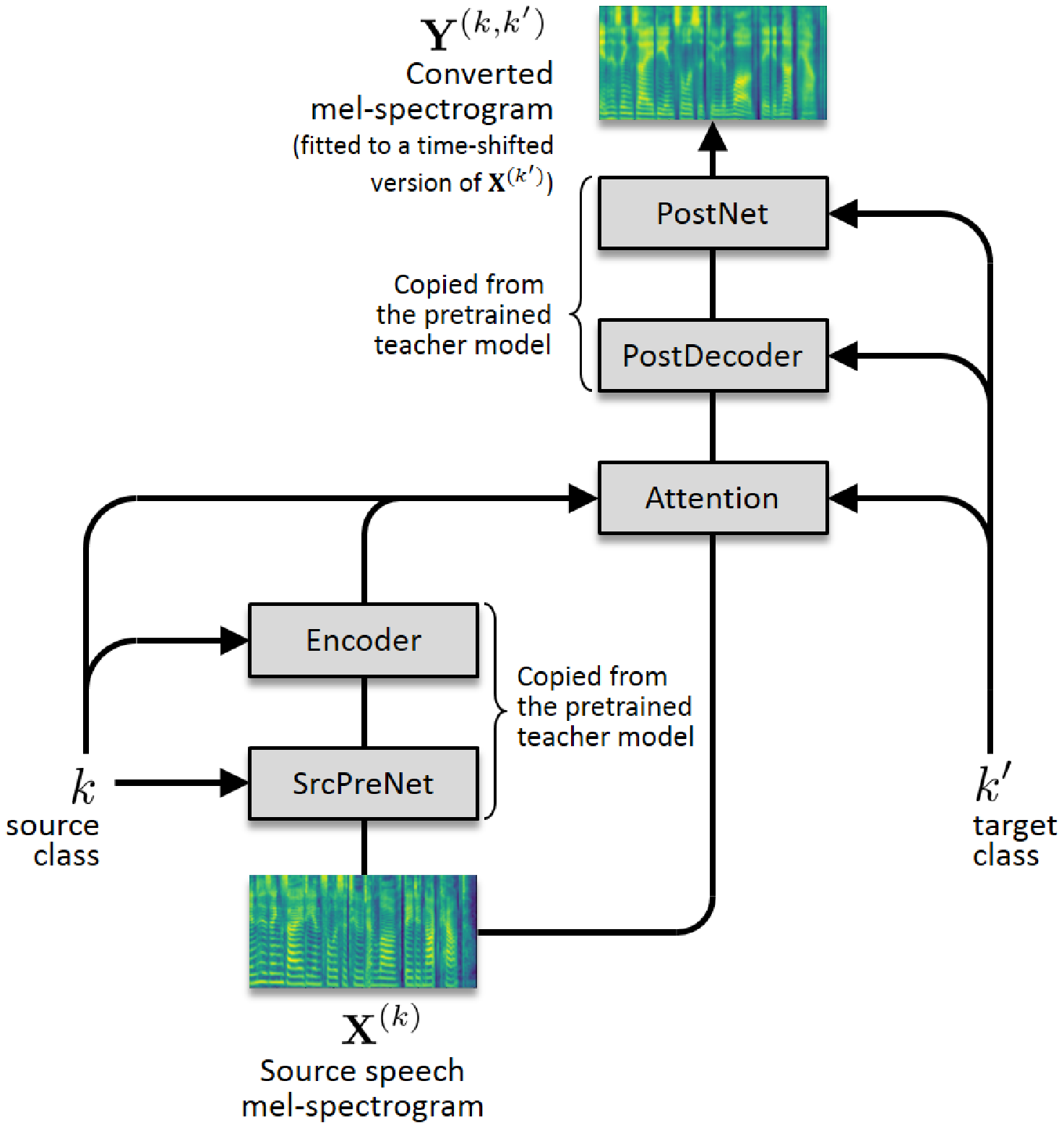}}
%\vspace{-1ex}
{\footnotesize (b) Student model}
\end{minipage}
\caption{Structures of the teacher and student models in FastS2S-VC. 
The teacher model is slightly modified from the one shown in \reffig{GeneralForm} so that the postdecoder does not receive direct input from the predecoder. The attention predictor in the student model is the only network that is trained during student training; the remaining networks are copied from the pretrained teacher model.}
\label{fig:TeacherStudent}
%\vspace{-1ex}
\end{figure*}

Note, however, that in this way, the speaking rate and rhythm of the converted speech will be exactly the same as that of input speech.
Here, we are concerned with a more challenging problem: allowing even the speaking rate and rhythm of input speech to be converted, under the constraint that the timings of the input and converted speech are synchronized at a certain interval.
If we choose not to skip performing attention matrix prediction at test time, we cannot avoid AR recursion as far as we use the above models as they are. 
Nevertheless, fast inference is still possible up to a point if we use sufficiently shallow architectures. 
However, making the network deeper to achieve better conversion quality comes at the cost of longer latency.
To achieve both good quality and short latency, we believe an NAR extension of S2S models is one of the keys.

\section{FastS2S-VC}

\subsection{Overall Structure}

Motivated by the above, 
the main idea we propose in this paper is an NAR extension of S2S models tailored to real-time VC. 
We call the VC method based on the proposed model ``FastS2S-VC''.

One major factor causing the AR structure in all the above models is the attention mechanism, which is designed to use both source and target mel-spectrograms to compute attention weight matrices. 
Therefore, at test time, the attention module will not be able to compute attention weights for a particular frame until the postnet has finished producing the target mel-spectrum in that frame.
To remove this restriction, the attention module must have a structure such that attention weight matrices are computable without target mel-spectrograms.
One possible solution to this would be to introduce an additional network that learns to predict the attention weight matrix only from a source mel-spectrogram along with source and target class labels. 
We call this network ``attention predictor''.
To facilitate training, we can adopt a teacher-student learning framework, in which one of the above AR-type models is used as the teacher model, as detailed below. 
Another factor causing the AR structure lies in the design such that the postdecoder takes the output of the predecoder as input. 
This restriction can be avoided simply by not allowing the postdecoder to use the predecoder output in its process.
The teacher model must also be redesigned accordingly.
To put the above together, the teacher and student models we study in this paper is structured as in \reffig{TeacherStudent}.
In the following, we discuss how the attention predictor should be designed and trained.
Note that since architectures containing recurrent units are not parallelizable, the architecture of the student model considered below assumes the use of the ConvS2S-VC2 or Transformer-VC2 architecture as the teacher model.

As detailed later, our real-time S2S-VC system operates in a sliding-window fashion designed so that the window length corresponds to the latency of the entire VC process. 
A straightforward implemention of this can cause problems when the speaking rate of the converted speech is very different from that of input speech.
For example, when the converted speech becomes faster, a silent segment must be inserted somewhere within each window, and when it becomes slower, some segment of the converted speech must be skipped. 
This may result in choppy-sounding or unintelligible speech even though the conversion process has been successful. 
To avoid this, the attention weight matrix should be represented in a form that is flexibly expandable and contractible to fit into a given window length. 
Furthermore, the peak of the attention weight distribution in each frame must be guaranteed to move only forward, so that some segments of the converted speech will not be skipped or repeated.

\subsection{Attention Predictor}

Given the above requirements and inspired by Graves' Gaussian attention \cite{Graves2013}, we propose using (unnormalized) Gaussian distribution functions to model attention weight matrices and designing the attention predictor to produce their parameters. 
The idea of using parametric functions such as Gaussian and logistic distribution functions to represent attention distributions for end-to-end TTS has already been proposed \cite{Vasquez2019,Battenberg2020}. 
However, the only point that our idea has in common with these previous studies is the use of parametric functions to represent attention distributions, but the way they are used is very different:
While the previous studies use parametric functions to represent each {\it column} of an attention weight matrix with the aforementioned AR structure in mind,
we use them to represent each {\it row} instead 
so that the prediction of each row of an attention weight matrix can be executed in parallel.
Another advantage is that this representation allows the length of the converted speech in each sliding window to be flexibly expandable or contractible to fit into a given window length without sacrificing the audio quality, thus facilitating smooth chunk-by-chunk conversion.

The attention module in the teacher model computes one attention weight matrix for the ConvS2S architecture and multiple attention weight matrices for the Transformer architecture, and uses it (them) to produce a time-warped version of the source context vector sequence $\Vec{Z}^{(k)}$. 
Let $\Vec{A}^{(k,k')}$ 
denote the attention weight matrix obtained by the attention module in the ConvS2S model
and $\Vec{A}_{1}^{(k,k')},\ldots,\Vec{A}_{H}^{(k,k')}$ 
denote the attention weight matrices obtained by the $L$th (final) attention module in the Transformer model, where $H$ is the number of heads in multi-head attention.
For the ConvS2S architecture, the output of the attention module is given by
\begin{align}
\Vec{R}^{(k,k')} = \Vec{V}^{(k)}\Vec{A}^{(k,k')},
\end{align}
where $\Vec{V}^{(k)}\in\mathbb{R}^{d/2 \times N}$ is one of the half splits of 
$\Vec{Z}^{(k)}\in\mathbb{R}^{d\times N}$ in the channel direction and $d$ is the dimension of each context vector.
For the Transformer architecture, 
the final attention module generates a time-warped context vector sequence
$\Vec{R}^{(k,k')}$,
given by
\begin{align}
\Vec{R}^{(k,k')} &= 
\Vec{W}
[\Vec{V}_{1}^{(k)}\Vec{A}_{1}^{(k,k')};
\ldots;
\Vec{V}_{H}^{(k)}\Vec{A}_{H}^{(k,k')}],
\end{align}
where $\Vec{V}_{1}^{(k)},\ldots,\Vec{V}_{H}^{(k)}\in \mathbb{R}^{d/H \times N}$ are linear projections of $\Vec{Z}^{(k)}\in\mathbb{R}^{d\times N}$ obtained using different learnable weight matrices, 
and $\Vec{W}\in\mathbb{R}^{d\times d}$ is another learnable weight matrix.

To design the student model, we now consider modeling each element $a_{h,n,m}$ of $\Vec{A}_h^{(k,k')}$ using a parametric function $\alpha_{h,n}(m)$ of the form 
\begin{align}
\alpha_{h,n}(m) = 
\phi_{h,n} 
\exp\bigg(-\frac{(m-\mu_{h,n})^2}{2\sigma_{h,n}^2}\bigg),
\label{eq:GaussAttention}
\end{align}
where $h$ is an index that can be dropped for the ConvS2S architecture and 
corresponds to one of the heads in multi-head attention for the Transformer architecture.
Obviously, the peak of $\alpha_{h,n}(m)$ is centered at $\mu_{h,n}$.
This representation can thus be interpreted to mean that time $n$ in the source speech is likely to correspond to time $\mu_{h,n}$ in the target speech. 
Since the time-ordering of $\mu_{h,1},\ldots,\mu_{h,N}$ should be non-decreasing from left to right, we would like them to satisfy $\mu_{h,1}\le \mu_{h,2} \le \cdots \le \mu_{h,N-1}\le \mu_{h,N}$. 
A convenient way to handle this constraint would be to reparametrize them using non-negative variables $\Delta_{h,1},\ldots,\Delta_{h,N}$ as 
\begin{align}
\mu_{h,1} &= \Delta_{h,1},
\\
\mu_{i,n} &= \mu_{i,n-1} + \Delta_{h,n}~~(n\ge 2).
\end{align}
Hence, 
the relationship between 
$\Vec{\mu}_{h}=[\mu_{h,1},\ldots,\mu_{h,N}]$
and 
$\Vec{\Delta}_{h}=[\Delta_{h,1},\ldots,\Delta_{h,N}]$
can be written in vector notation as
\begin{align}
\underbrace{
[\mu_{h,1},\ldots,\mu_{h,N}]
}_{\Vec{\mu}_{h}}
= 
\underbrace{
[\Delta_{h,1},\ldots,\Delta_{h,N}]
}_{\Vec{\Delta}_{h}}
\underbrace{
\begin{bmatrix}
1&\cdots&1\\
 &\ddots&\vdots\\
0&      &1
\end{bmatrix}
}_{\Vec{U}}.
\label{eq:mu_and_delta}
\end{align}
\begin{comment}
\begin{align}
\underbrace{
\begin{bmatrix}
\mu_{h,1}\\
\vdots\\
\mu_{h,N}
\end{bmatrix}
}_{\Vec{\mu}_{h}}
= 
\underbrace{
\begin{bmatrix}
1&&0\\
\vdots&\ddots&\\
1&\cdots&1
\end{bmatrix}
}_{\Vec{U}}
\underbrace{
\begin{bmatrix}
\Delta_{h,1}\\
\vdots\\
\Delta_{h,N}
\end{bmatrix}
}_{\Vec{\Delta}_{h}}.
\end{align}
\end{comment}
We design the attention predictor in the student model as a network that takes the source context vector sequence $\Vec{Z}^{(k)}$ and source and target class indices $(k,k')$, and produces an array $\Vec{\theta}^{(k,k')}$ consisting of the parameters of the Gaussians, 
$\Vec{\theta}^{(k,k')} = [\Vec{\Delta}; \Vec{\Sigma}; \Vec{\Phi}]\in \mathbb{R}^{3H \times N}$, where 
$\Vec{\Delta}=(\Delta_{h,n})_{H\times N}$, 
$\Vec{\Sigma}=(\sigma_{h,n})_{H\times N}$, and 
$\Vec{\Phi}=(\phi_{h,n})_{H\times N}$
($H=1$ for the ConvS2S architecture and $H\ge 1$ for the Transformer architecture),
as an intermediate output.
Note that here we have used $[;]$ to denote vertical concatenation of arrays, matrices or vectors with compatible sizes.
Once $\Vec{\theta}^{(k,k')}$ is determined, the attention predictor computes attention distribution functions $\Vec{\alpha}_h^{(k,k')}=(\alpha_{h,n}(m))_{N\times M}$ $(h=1,\ldots,H)$ using \refeqs{mu_and_delta}{GaussAttention}.
Since attention distributions are assumed to sum to 1 along the $n$-axis, 
each element of $\Vec{\alpha}_h^{(k,k')}$ is further normalized as
\begin{align}
\alpha_{h,n}(m)\leftarrow \frac{\alpha_{h,n}(m)}{\sum_{n'} \alpha_{h,n'}(m)}.
\label{eq:normalize}
\end{align}
The attention module in the student model finally produces a time-warped version of $\Vec{Z}^{(k)}$ in the same manner as in the teacher model, by using $\Vec{\alpha}_h^{(k,k')}$ instead of $\Vec{A}_h^{(k,k')}$.  

\begin{figure}[t!]
\centering
\begin{minipage}[t]{.6\linewidth}
  \centerline{\includegraphics[width=.98\linewidth]{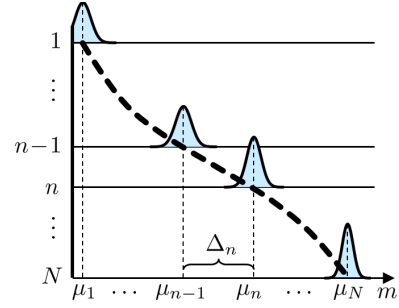}}
  %\vspace{-1ex}
  \caption{Attention distribution function $\alpha_n(m)$.}
\label{fig:AttentionApproximator}
\end{minipage}
%\vspace{-1ex}
\end{figure}

\begin{figure}[t!]
\centering
\begin{minipage}[t]{.49\linewidth}
\centering
\centerline{\includegraphics[height=5.4cm]{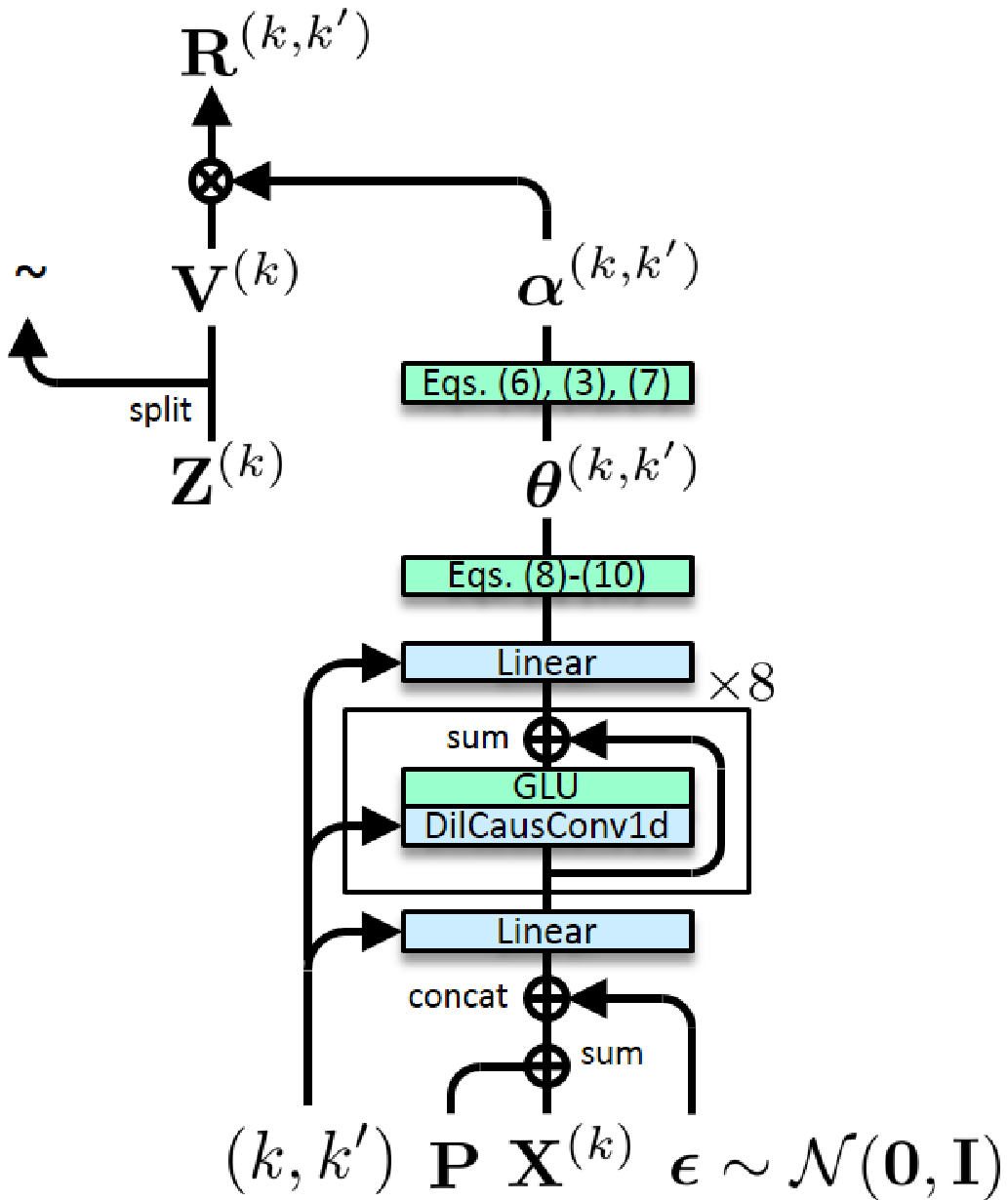}}
%\vspace{-1ex}
{\footnotesize (a) ConvS2S}
\end{minipage}
\begin{minipage}[t]{.49\linewidth}
\centering
\centerline{\includegraphics[height=5.4cm]{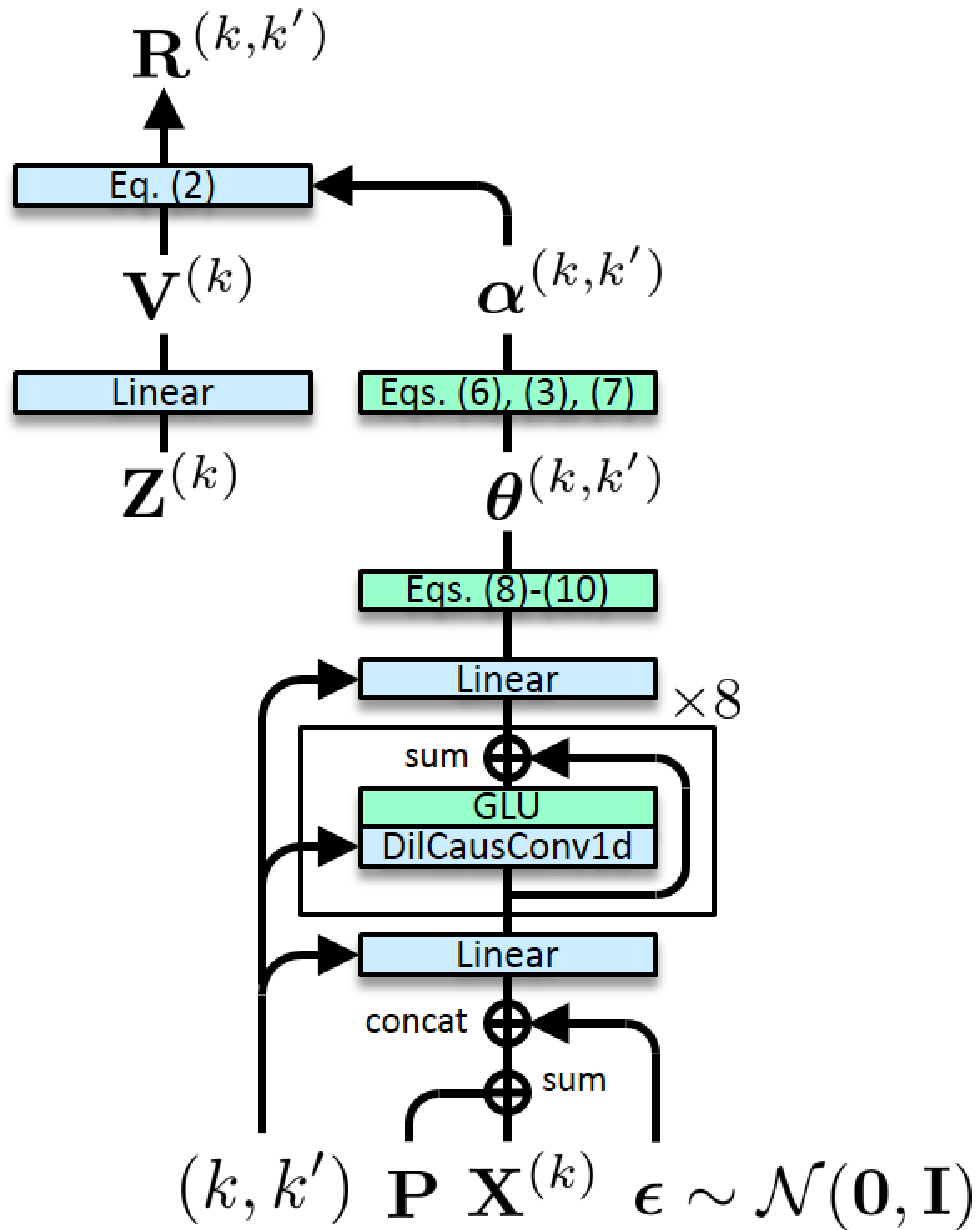}}
%\vspace{-1ex}
{\footnotesize (b) Transformer}
\end{minipage}
\caption{Architectures of the student attention module assuming the use of the (a) ConvS2S and (b) Transformer architectures as the teacher model. In (b), $\Vec{\alpha}^{(k,k')}$ and $\Vec{V}^{(k)}$ are abbreviations of the sets $\{\Vec{\alpha}_{h}^{(k,k')}\}_{h}$ and $\{\Vec{V}_{h}^{(k)}\}_{h}$, respectively. }
\label{fig:AttentionPredictor}
%\vspace{-1ex}
\end{figure}

Since all the parameters of the Gaussians must be non-negative, we include a layer that performs
\begin{align}
\Vec{\Delta}&\leftarrow {\rm absolute}(\Vec{\Delta}),\\
\Vec{\Sigma}&\leftarrow \min(\max({\rm absolute}(\Vec{\Sigma}),0.001),1.0),\label{eq:sigma_update}\\ 
\Vec{\Phi}&\leftarrow 0.2{\rm sigmoid}(\Vec{\Phi})+0.8,\label{eq:phi_update}
\end{align}
before producing $\Vec{\theta}^{(k,k')}$, where ${\rm absolute}(\cdot)$ and ${\rm sigmoid}(\cdot)$ denote elementwise absolute and sigmoid functions, and $\min(a,b)$ and $\max(a,b)$ denote functions that return the elementwise minimum and maximum of $a$ and $b$. 
Note that \refeqs{sigma_update}{phi_update} are to prevent each Gaussian from becoming too wide and  vanishing, respectively. 

The operation in the attention predictor network must be parallelizable and causal. 
We thus choose to design the network with a fully convolutional architecture, consisting of two fully-connected linear layers and eight dilated causal convolution layers with the kernel sizes of 5 and the dilation factors of 1, 3, 9, 27, 1, 3, 9, and 27, respectively, each followed by a GLU. 

\subsection{Random Noise Input}

Speech fluctuates in time from utterance to utterance.
Therefore, the timing differences between source and target speech also fluctuate, and the rules governing these fluctuations are neither unique nor deterministic. 
In other words, the distribution of the time-warping functions the attention module needs to predict is likely to be multimodal. 
Thus, the problem the attention module must handle is a one-to-many mapping problem. 
Although AR models are generally known to be reasonably good at handling one-to-many mapping problems, since the attention predictor does not rely on an AR structure, it must have some mechanism that can deal with this problem nicely.
Although very simple, we propose including randomly drawn samples in the input to the attention predictor in the hope that these samples will be used to explain the randomness related to the temporal fluctuations in the timing differences between source and target speech that cannot be explained by deterministic rules.
This simple idea actually worked out well.

To put the above together, the architecture of the student attention module can be configured as in \reffig{AttentionPredictor}.

\subsection{Student Training Objective}

In the student training phase, the source prenet, encoder, postdecoder, and postnet are assumed to be fixed at the ones copied from a pretrained teacher model, and the only network to be trained is the attention predictor.
For details about the teacher training objectives, please refer to \cite{Kameoka2018IEEE-TASLP_ConvS2S-VC,Kameoka2020IEEE-TASLP_VTN}.

Let $\Vec{X}^{(k)}\in\mathbb{R}^{D\times N}$ and $\Vec{X}^{(k')}\in\mathbb{R}^{D\times M}$ be the mel-spectrograms of a pair of parallel utterances, where $k$ and $k'$ denote the source and target classes. 
As a start-of-sequence token, an all-zero vector is appended to the left end of the target sequence only, namely, $\Vec{X}^{(k')}\leftarrow [\Vec{0},\Vec{X}^{(k')}]$. This results in a length of $M+1$. 
Given this single pair of utterances,
the main loss to be minimized is the mean absolute error between the model output $\Vec{Y}^{(k,k')}$ and a time-shifted version of the target mel-spectrogram $\Vec{X}^{(k')}$, as in the teacher training:
\begin{align}
\mathcal{L}_{0} = 
{\textstyle
\frac{1}{M}
}
\| \Vec{Y}_{1:M}^{(k,k')} - \Vec{X}_{2:M+1}^{(k')} \|_1,
\end{align}
where 
the notation $\Vec{X}_{i:j}$ refers to the submatrix consisting of the $i$th to $j$th columns of the matrix $\Vec{X}$.
In addition to this, we can use the attention weight matrix $\Vec{A}_h^{(k,k')}$ computed by the teacher model for each source-target utterance pair as a regression target to guide the attention predictor output $\Vec{\alpha}_h^{(k,k')}$. 
However, it transpired that using the difference between $\Vec{A}_h^{(k,k')}$ and $\Vec{\alpha}_h^{(k,k')}$ alone as the loss did not work very well. This was because both the regression target and approximator were likely to be very sparse: Even when the peaks of the regression target and approximator were slightly apart, the gradients tended to vanish easily. 
For faster and more stable training,
we have found it useful to use 
\begin{align}
\mathcal{L}_{1} = 
{\textstyle
\frac{1}{HN}
\sum_{h,n}
} 
(|\mu_{h,n} - \hat{\mu}_{h,n}| + |\sigma_{h,n} - \hat{\sigma}_{h,n}|),
\end{align}
as an auxiliary loss, 
where $\hat{\mu}_{h,n}$ and $\hat{\sigma}_{h,n}$ are the mean and standard deviation of time $m$ when each row $a_{h,n,1}^{(k,k')},\ldots,a_{h,n,M}^{(k,k')}$ of $\Vec{A}_h^{(k,k')}$ is seen as a histogram of $m=1,\ldots,M$, given by
\begin{align}
\hat{\mu}_{h,n} &= \frac{\sum_{m=1}^{M} a_{h,n,m}^{(k,k')} m}{\sum_{m=1}^{M} a_{h,n,m}^{(k,k')}},\\
\hat{\sigma}_{h,n}^2 &= \frac{\sum_{m=1}^{M} a_{h,n,m}^{(k,k')} (m - \hat{\mu}_{h,n})^2}{\sum_{m=1}^{M} a_{h,n,m}^{(k,k')}}.
\end{align}
To further enforce $\Vec{\alpha}_{h}^{(k,k')}$ to be as diagonally dominant and orthogonal as possible, we introduce the diagonal \cite{Tachibana2018short} and orthogonal \cite{Kameoka2018IEEE-TASLP_ConvS2S-VC} attention losses:
\begin{align}
\mathcal{L}_{2} &= 
{\textstyle
\frac{1}{HNM}\sum_{h}
}
\| \Vec{G}_{N\times M}(\nu) \odot \Vec{\alpha}_h^{(k,k')} \|_1,\\
\mathcal{L}_{3} &= 
{\textstyle
\frac{1}{HN^2}\sum_{h}
}
\| \Vec{G}_{N\times N}(\rho) \odot (\Vec{\alpha}_h^{(k,k')}\Vec{\alpha}_h^{(k,k')}{}^{\mathsf T}) \|_1,
\end{align}
where $\odot$ denotes an elementwise product, and 
$\Vec{G}_{N\times M}(\nu) \in\mathbb{R}^{N \times M}$ 
is a non-negative weight matrix  
whose $(n,m)$th element $g_{n,m}$ is defined as 
$g_{n,m} = 1- e^{-(n/N-m/M)^2/2\nu^2}$.

Given all examples of parallel utterances, 
the total training loss for the student model to be minimized is given as
\begin{align}
\mathcal{L} = 
\sum_{k,k'}
\mathbb{E}_{\Vec{X}^{(k)},\Vec{X}^{(k')}}
\left\{
\mathcal{L}_{0}
+
\lambda_{1}
\mathcal{L}_{1}
+
\lambda_{2}
\mathcal{L}_{2}
+
\lambda_{3}
\mathcal{L}_{3}
\right\},
\end{align}
where 
$\mathbb{E}_{\Vec{X}^{(k)},\Vec{X}^{(k')}}\{\cdot\}$ is 
the sample mean over all the training examples 
of parallel utterances of classes $k$ and $k'$, 
and 
$\lambda_{1}\ge 0$, $\lambda_{2}\ge 0$ and $\lambda_{3}\ge 0$
are regularization parameters.

\subsection{Conversion process}
\label{subsec:conversion}

Once the attention predictor has been trained, the conversion process simply becomes a cascade of the five trained networks, namely the source prenet, encoder, attention module, postdecoder, and postnet. 
It is worth noting that this process can be executed in parallel without recursion. 
After converting the mel-spectrogram of input speech through this process, we finally generate a waveform using Parallel WaveGAN \cite{Yamamoto2020}.

\subsection{Real-Time Implementation Details}
\label{subsec:FastS2S-VC_RT}

\begin{figure}[t!]
\centering
\begin{minipage}[t]{.85\linewidth}
%\centering
\centerline{\includegraphics[width=.75\linewidth]{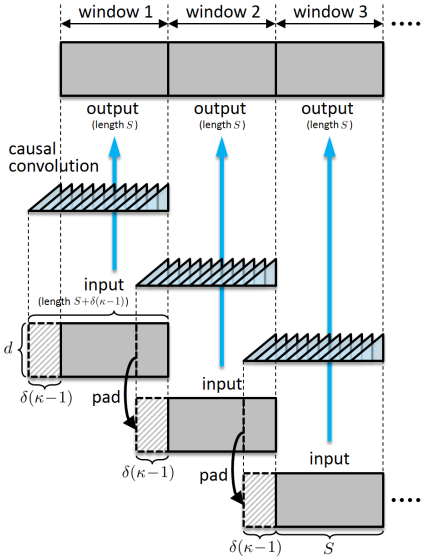}}
\caption{Implementation of sliding-window causal convolutions. }
\label{fig:SlidingWindow}
\end{minipage}
%\vspace{-1ex}
\end{figure}

\begin{figure}[t!]
\centering
\begin{minipage}[t]{.85\linewidth}
%\centering
\centerline{\includegraphics[width=.6\linewidth]{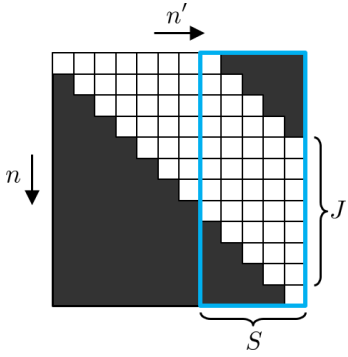}}
\caption{Image of a mask by which a self-attention matrix is to be multiplied elementwise. 
The black and white areas represent the 0 and 1 elements, respectively. Only the submatrix of a self-attention matrix corresponding to the area surrounded by the blue line is used to compute the output of a self-attention layer.}
\label{fig:Mask}
\end{minipage}
%\vspace{-1ex}
\end{figure}

Here, we desribe some of the techniques to implement real-time systems of FastS2S-VC.  
First, all the convolution and self-attention layers must be causal.
To make the best use of the parallelizable structures of the convolution and self-attention layers, we take a sliding-window approach.
Fortunately, causal convolutions can be computed in a sliding-window fashion without any approximation, as explained in \reffig{SlidingWindow}. 
Namely, consider dividing an input sequence into non-overlapping chunks and performing convolution on each chunk in a way that a portion of the previous chunk is padded to the left end of the current chunk. 
When the padding size is $\delta(\kappa-1)$, where 
$\delta$ and $\kappa$ are the dilation factor and kernel size of the convolution, respectively,
the concatenation of all the outputs becomes exactly the same as the output of that layer where the entire sequence is given as input.
Note that since the architecture of Parallel WaveGAN consists of non-causal convolution layers, we have redesigned all the convolution layers to be causal so that the same technique can be used.
As for the self-attention layers, 
the distance between elements in a sequence that can interact with each other must be limited, 
since 
storing all the elements and keeping track of their interdependencies are not possible in real-time streaming scenarios.
To avoid having to compute the self-attention between two elements that are more than a certain distance $J$ apart, we mask each self-attention matrix so that position $n$ can depend only on the elements at positions $n-1,\ldots,n-J$. This is equivalent to multiplying a binary matrix like the one in \reffig{Mask} elementwise by an unconstrained self-attention matrix. 
This constraint makes the training and runtime conditions consistent,
and allows us to compute the process of a self-attention layer in a sliding-window fashion without approximation by padding a portion of length $J$ of the previous chunk of an input sequence to the left end of the current chunk, and passing a sequence consisting of only the last $S$ elements of the ouput sequence to the next layer.

One of the key advantages of representing each row of an attention matrix as a Gaussian distribution function in FastS2S-VC
is the flexibility to expand or contract the length of the converted version of each chunk of input speech to match the window length $S$. 
If we use $\bar{\mu}_1$ and $\bar{\mu}_S$ to denote
the means of 
$\mu_{1,1},\ldots,\mu_{H,1}$ and 
$\mu_{1,S},\ldots,\mu_{H,S}$, respectively, 
one simple way would be to transform the centers $\mu_{h,1},\ldots,\mu_{h,S}$ of the Gaussians linearly:
\begin{align}
\mu_{h,n} \leftarrow
\frac{(S-1)(\mu_{h,n}-\bar{\mu}_1)}{\bar{\mu}_S - \bar{\mu}_1} + 1~(n=1,\ldots,S),
\end{align}
so that 
$\bar{\mu}_1 = 1$ and $\bar{\mu}_S = S$,
and compute 
$\Vec{\alpha}_h^{(k,k')}$ and $\Vec{R}^{(k,k')}$ accordingly. 
By keeping $\sigma_{h,1},\ldots,\sigma_{h,S}$ unchanged, the resulting $\Vec{\alpha}_h^{(k,k')}$ can be made neither too blurry nor too sharp.

\section{Experiments}
\label{sec:experiments}

\begin{table*}[t!]
\centering
\caption{Methods for comparison}
\label{tab:methods}
\begin{adjustbox}{center}
\begin{tabular}{l|l|l V{3} c}
\thline
\multicolumn{3}{c V{3}}{Method}&\multirow{2}{*}{Explanation}\\\cline{1-3}
Base model&Processing&Architecture& \\\thline
\multirow{1}{*}{S2S-VC}&BAT&
RNN/Conv/Trans&
\parbox{.6\linewidth}{
\medskip
Batch processing (BAT) version of S2S-VC 
with different architectures 
(RNNS2S-VC2, ConvS2S-VC2, and Transformer-VC2).
Conversion is performed using AR recursion.
\medskip
}
\\\hline
\multirow{2}{*}{FastS2S-VC}&BAT&Conv/Trans&
\parbox{.6\linewidth}{
\medskip
Batch processing (BAT) version of FastS2S-VC. 
All the networks are designed to be causal.
Conversion is done in parallel without AR recursion.
\medskip
}
\\\cline{2-4}
&RT&Conv/Trans&
\parbox{.6\linewidth}{
\medskip
Real-time processing (RT) version of FastS2S-VC described in \refsubsec{FastS2S-VC_RT}.
All the networks are designed to be causal.
Conversion is done in a sliding-window fashion without AR recursion.
The function to convert the local speaking rate (rhythm) of input speech is enabled.
\medskip
}
\\\thline
\end{tabular}
\end{adjustbox}
\end{table*}

\subsection{Experimental Settings}
\label{subsec:expcond}

To evaluate the conversion quality of the proposed methods, 
we conducted objective and subjective evaluation experiments involving
speaker-identity and emotional-expression conversion tasks. 
For the speaker-identity conversion task, we used the CMU Arctic database \cite{Kominek2004short},
which consists of recordings of 1,132 phonetically balanced English utterances  
spoken by four US English speakers, 
clb (female), 
bdl (male),
slt (female), and 
rms (male).
Therefore, there were a total of 12 combinations of source and target speakers.
For each speaker, the first 1,000 utterances were used as 
the training set, and the remaining 132 utterances were used as the evaluation set.
For the emotional-expression conversion task, we used audio signals of 503 phonetically balanced sentences from the ATR Japanese Speech Database \cite{Kurematsu1990}, read by one voice actress with four different emotional expressions (neutral, angry, sad, and happy).
For each expression, the utterances corresponding to the first
450 sentences were used as the training set, and those corresponding to the remaining 53 sentences were used as the evaluation set.
All the speech signals were sampled at 16 kHz. 
The mel-spectrogram with 80 frequency bands of each utterance was computed with a frame length of 64 ms and a hop size of 8 ms. 
The reduction factor $r$ was set to 4.
Hence, the dimension of the acoustic feature was $D=80\times 4=320$.

As described in \refsubsec{GeneralForm}, we used Parallel WaveGAN \cite{Yamamoto2020} for waveform generation from mel-spectrograms.
For its implementation, we used the unofficial source code available on github\footnote{https://github.com/kan-bayashi/ParallelWaveGAN}. Specifically, we trained the speaker-independent version on the same training set used to train the core feature mapping models.
All algorithms were implemented in PyTorch and run on a single Tesla V100 SXM2 GPU with a 32.0 GB memory
and an Intel(R) Xeon(R) Gold 5218 16-core CPU @ 2.30GHz.

\subsection{Methods for Comparison}

The methods used for comparison in the current experiments are listed in \reftab{methods}.
These methods are categorized by base model, processing type, and architecture type,
where
the base model includes the baseline S2S-VC model and the proposed FastS2S-VC model,
the processing type includes batch (BAT) and real-time (RT) processing, and 
the architecture type includes recurrent (RNN), convolutional, and Transformer architectures. 
Note that 
existing S2S model-based VC methods \cite{Zhang2018short,Biadsy2019short}, including our RNNS2S-VC2, fall into the category of ``S2S-VC--BAT--RNN'', albeit with some differences in architectural details and training objectives.

We also chose the open-source VC system 
called sprocket \cite{Kobayashi2018short} as a baseline in the subjective listening tests. 
To run this method, we used the source code provided by its author\footnote{https://github.com/k2kobayashi/sprocket}.
Note that this system was used as a baseline system in the
Voice Conversion Challenge (VCC) 2018 \cite{Lorenzo-Trueba2018short}.

\begin{figure}[t!]
\centering
\begin{minipage}[t]{.48\linewidth}
\centering
\centerline{\includegraphics[width=.98\linewidth]{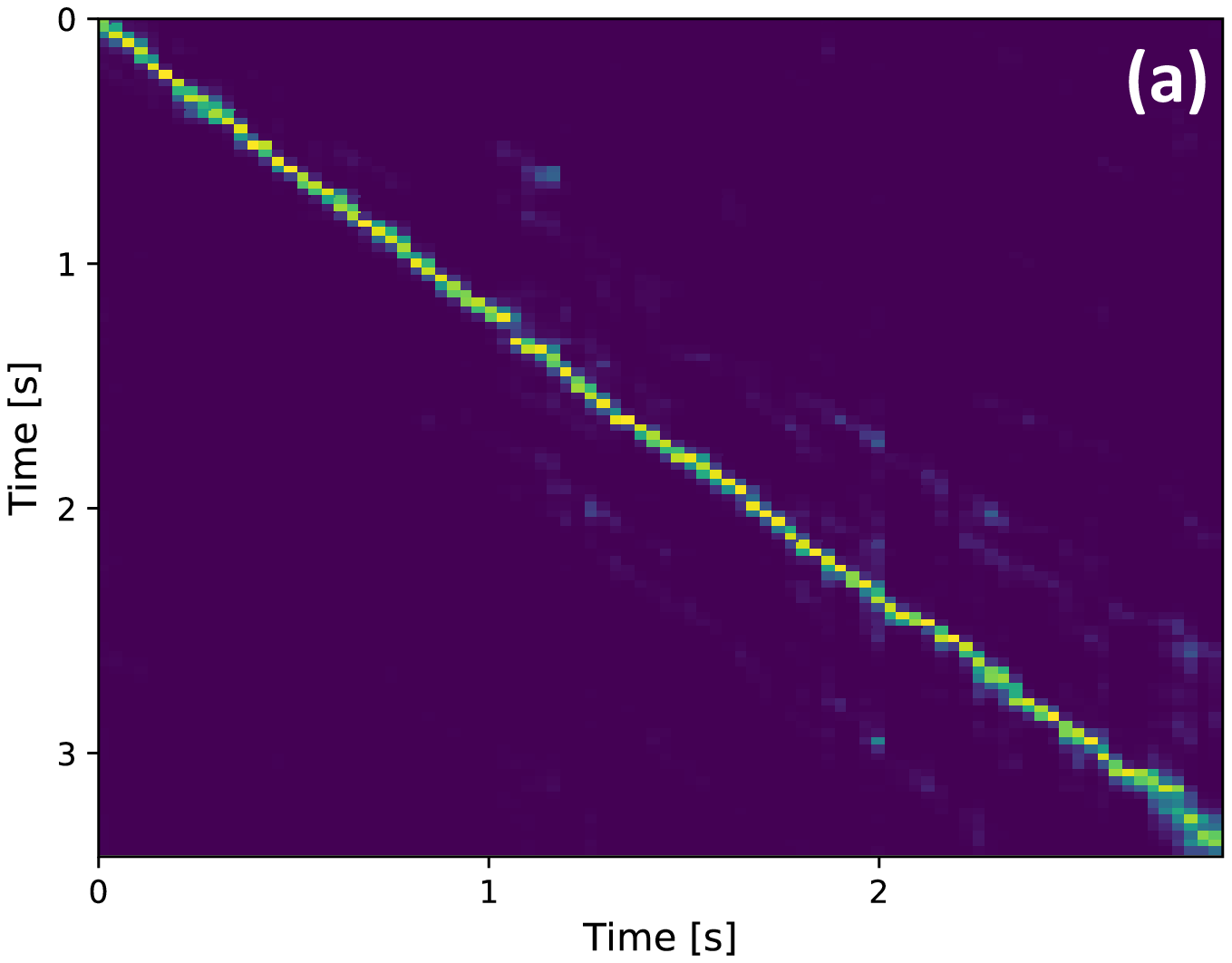}}
\end{minipage}
\\
\medskip
\begin{minipage}[t]{.48\linewidth}
\centering
\centerline{\includegraphics[width=.98\linewidth]{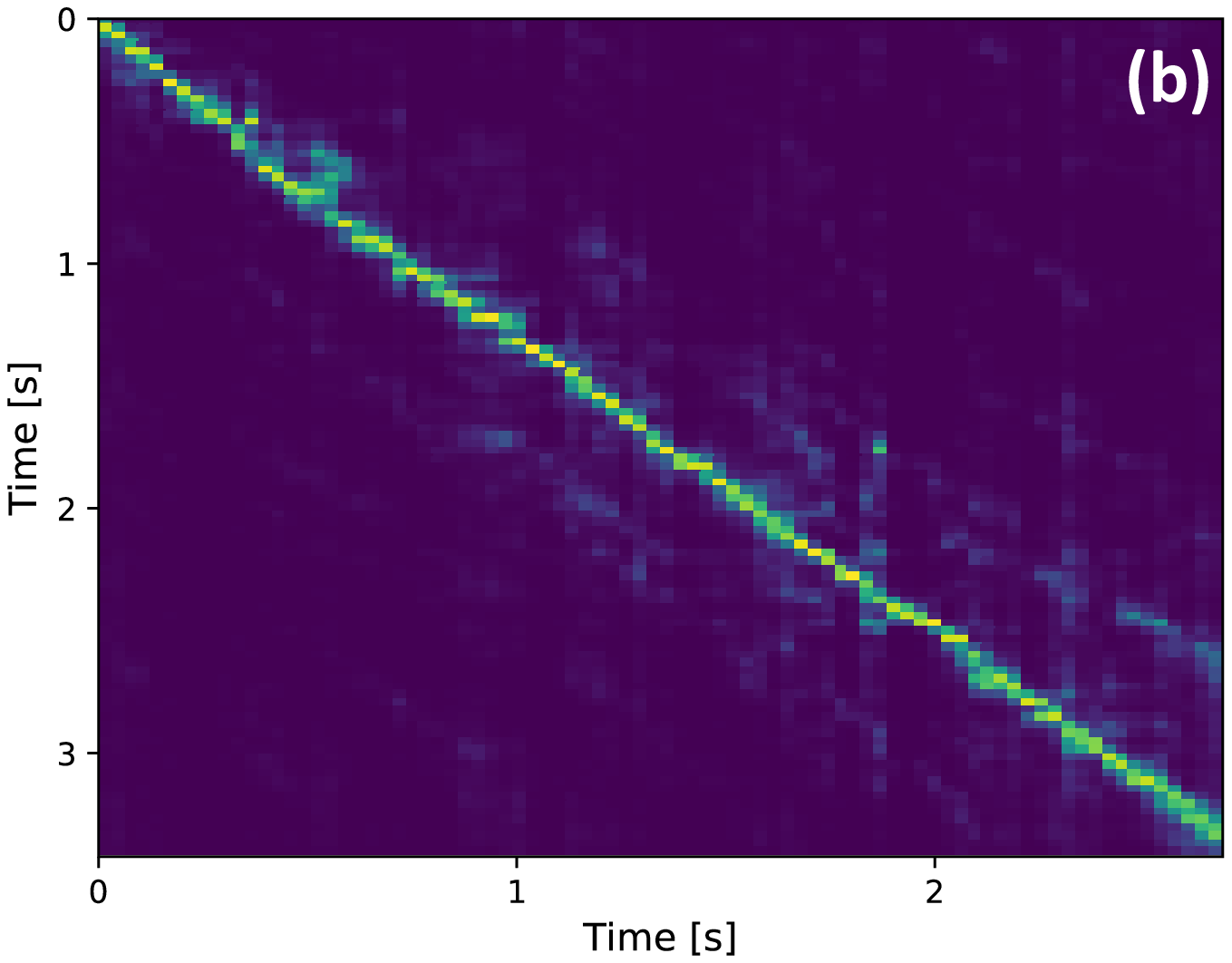}}
\end{minipage}
\begin{minipage}[t]{.48\linewidth}
\centering
\centerline{\includegraphics[width=.98\linewidth]{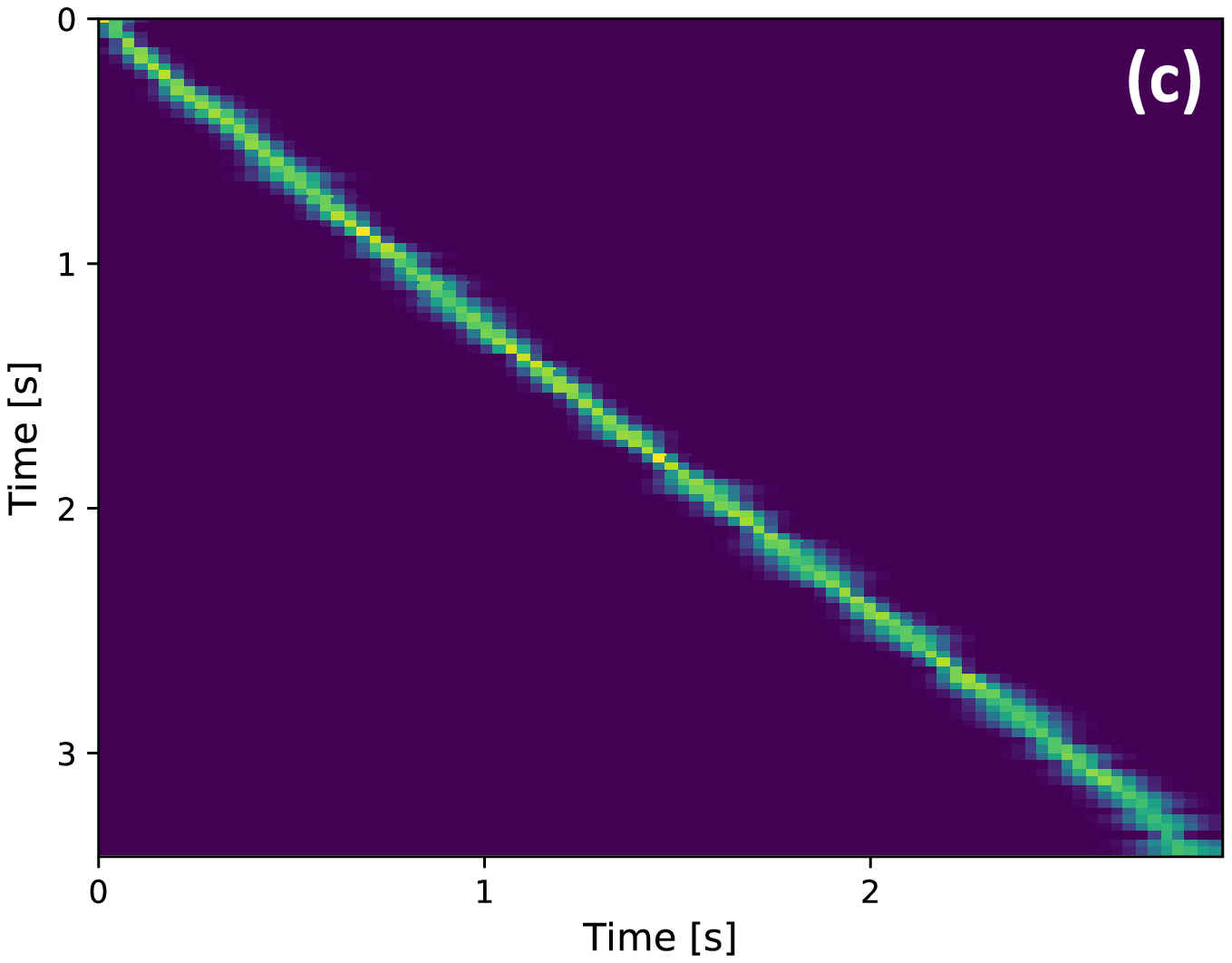}}
\end{minipage}
\\
\medskip
\centering
\begin{minipage}[t]{.98\linewidth}
\centering
\centerline{\includegraphics[width=.98\linewidth]{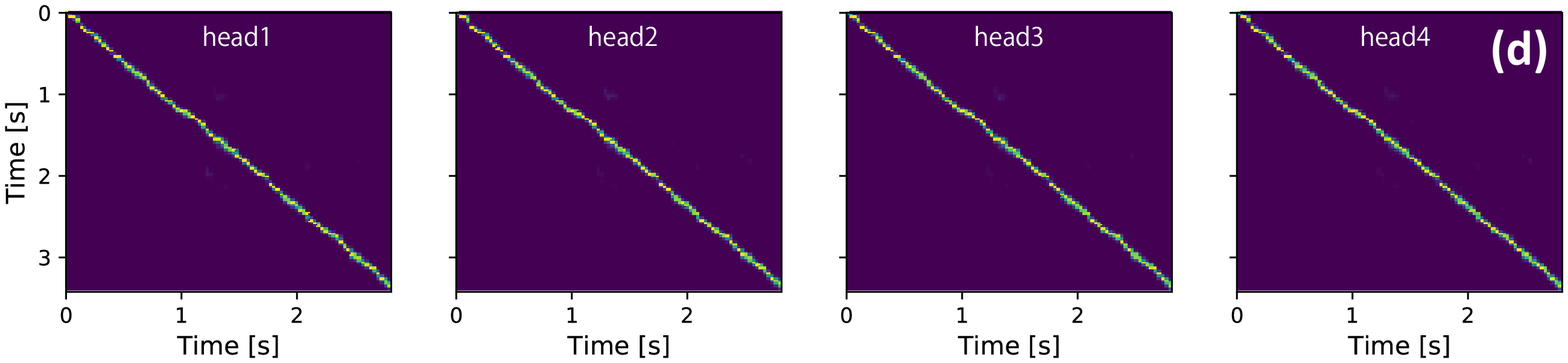}}
\end{minipage}
\\
\medskip
\begin{minipage}[t]{.98\linewidth}
\centering
\centerline{\includegraphics[width=.98\linewidth]{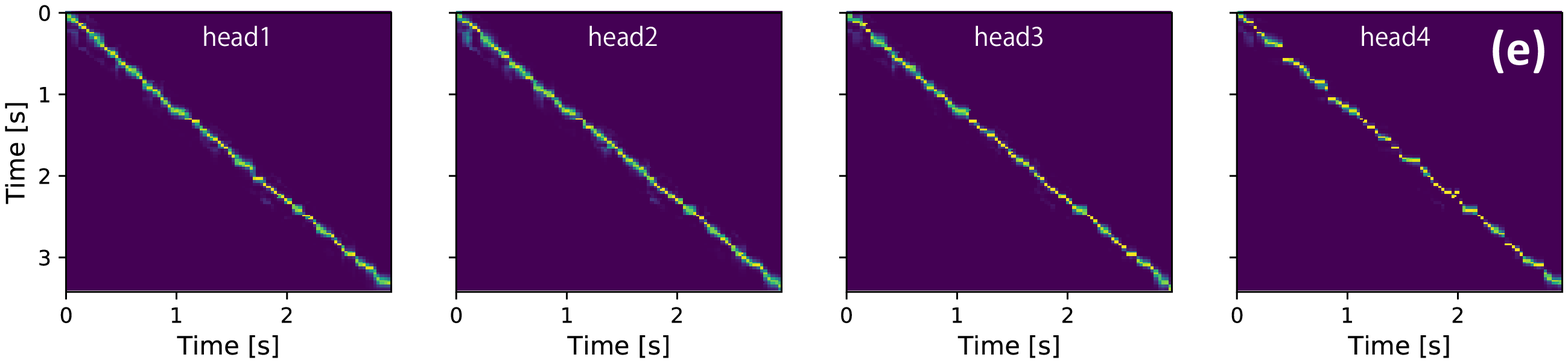}}
\end{minipage}
\\
\medskip
\begin{minipage}[t]{.98\linewidth}
\centering
\centerline{\includegraphics[width=.98\linewidth]{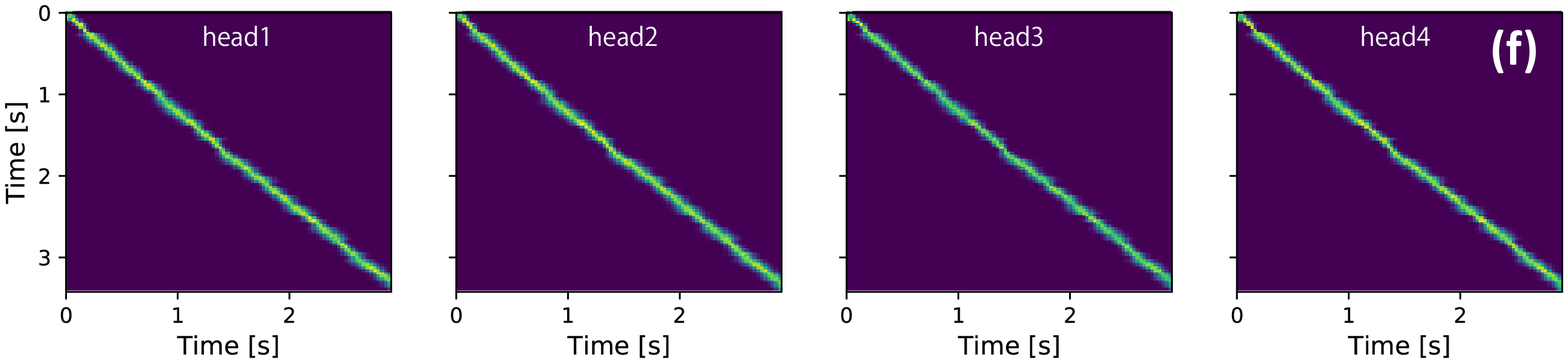}}
\end{minipage}
\caption{Attention matrices predicted by 
(a) S2S-VC--BAT--Conv,
the (b) teacher and (c) student models in FastS2S-VC--BAT--Conv,
(d) S2S-VC--BAT--Trans, and
the (e) teacher and (f) student models in FastS2S-VC--BAT--Trans
from 
the same utterance of clb when the target speaker was bdl.
In (d) and (e),
only the attention matrices produced from the last decoder layer are
shown.
}
\label{fig:arctic_b0411}
%\vspace{-1ex}
\end{figure}

\begin{table*}[t!]
\centering
\caption{MCDs and LFCs obtained with S2S-VC (BAT) and FastS2S-VC (BAT/RT).
}
\label{tab:mcd_baseline_comp}
(a) MCD\\
\begin{tabular}{l | l V{3} c|c|c|c|c|c|c}
\thline
\multicolumn{2}{c V{3}}{Speakers}&\multicolumn{3}{c|}{S2S-VC (BAT)}
&\multicolumn{2}{c|}{FastS2S-VC (BAT)}&\multicolumn{2}{c}{FastS2S-VC (RT)}
\\\cline{1-2}\cline{3-9}
source&target&RNN&Conv&Trans&Conv&Trans&Conv&Trans\\\thline
      &   bdl&$6.54$&$6.44$&$6.70$&$6.45$&$6.63$&$6.44$&$6.55$\\
   clb&   slt&$6.19$&$5.80$&$6.11$&$5.95$&$6.22$&$5.84$&$6.18$\\
      &   rms&$6.59$&$6.14$&$6.45$&$6.25$&$6.52$&$6.26$&$6.53$\\\hline
      &   clb&$6.30$&$5.89$&$6.24$&$6.09$&$6.47$&$6.17$&$6.48$\\
   bdl&   slt&$6.34$&$5.87$&$6.25$&$6.08$&$6.37$&$6.14$&$6.32$\\
      &   rms&$6.78$&$6.19$&$6.50$&$6.44$&$6.60$&$6.84$&$7.15$\\\hline
      &   clb&$6.27$&$5.86$&$6.13$&$6.03$&$6.32$&$6.11$&$6.30$\\
   slt&   bdl&$6.66$&$6.47$&$6.70$&$6.56$&$6.79$&$6.56$&$6.63$\\
      &   rms&$6.66$&$6.14$&$6.48$&$6.35$&$6.58$&$6.52$&$6.80$\\\hline
      &   clb&$6.31$&$6.00$&$6.42$&$6.14$&$6.44$&$6.23$&$6.41$\\
   rms&   bdl&$6.59$&$6.51$&$6.74$&$6.56$&$6.82$&$6.54$&$6.65$\\
      &   slt&$6.40$&$5.90$&$6.28$&$6.10$&$6.43$&$5.98$&$6.28$\\\hline
\multicolumn{2}{c V{3}}{All pairs}
             &$6.47$&$6.10$&$6.41$&$6.25$&$6.52$&$6.28$&$6.50$\\\thline   
\end{tabular}
\\
\medskip
(b) LFC\\
\begin{tabular}{l | l V{3} c|c|c|c|c|c|c}
\thline
\multicolumn{2}{c V{3}}{Speakers}&\multicolumn{3}{c|}{S2S-VC (BAT)}
&\multicolumn{2}{c|}{FastS2S-VC (BAT)}&\multicolumn{2}{c}{FastS2S-VC (RT)}
\\\cline{1-2}\cline{3-9}
source&target&RNN&Conv&Trans&Conv&Trans&Conv&Trans\\\thline
      &   bdl&$0.621$&$0.773$&$0.798$&$0.803$&$0.731$&$0.772$&$0.688$\\
   clb&   slt&$0.657$&$0.841$&$0.817$&$0.883$&$0.890$&$0.851$&$0.886$\\
      &   rms&$0.515$&$0.738$&$0.529$&$0.699$&$0.642$&$0.663$&$0.610$\\\hline
      &   clb&$0.676$&$0.834$&$0.732$&$0.793$&$0.751$&$0.751$&$0.771$\\
   bdl&   slt&$0.729$&$0.840$&$0.795$&$0.824$&$0.849$&$0.825$&$0.827$\\
      &   rms&$0.489$&$0.655$&$0.493$&$0.666$&$0.600$&$0.683$&$0.517$\\\hline
      &   clb&$0.692$&$0.832$&$0.805$&$0.811$&$0.823$&$0.799$&$0.853$\\
   slt&   bdl&$0.636$&$0.757$&$0.752$&$0.805$&$0.717$&$0.726$&$0.687$\\
      &   rms&$0.497$&$0.731$&$0.640$&$0.669$&$0.705$&$0.714$&$0.554$\\\hline
      &   clb&$0.692$&$0.731$&$0.782$&$0.785$&$0.764$&$0.704$&$0.721$\\
   rms&   bdl&$0.597$&$0.757$&$0.690$&$0.774$&$0.717$&$0.737$&$0.582$\\
      &   slt&$0.741$&$0.816$&$0.778$&$0.831$&$0.787$&$0.764$&$0.781$\\\hline
\multicolumn{2}{c V{3}}{All pairs}
             &$0.626$&$0.772$&$0.732$&$0.780$&$0.747$&$0.750$&$0.732$\\\thline   
\end{tabular}
\end{table*}

\subsection{Hyperparameter Settings in Model Training}

$\lambda_1$, $\lambda_2$, and $\lambda_3$
were set at 1, 2000, and 2000, respectively.
Both $\nu$ and $\rho$ were set at 0.3.

%All the networks were trained simultaneously with random initialization.
Adam optimization \cite{Kingma2015short} was used for model training where 
the mini-batch size was 16 for all the models.
70,000 iterations were run for the baseline S2S-VC models and the teacher models in FastS2S-VC, 
and 300,000 iterations for the student models. 
The learning rate and the exponential decay rate
for the first moment 
for Adam were set at $5\times 10^{-5}$ and 0.9.

\subsection{Objective Performance Measures}

The evaluation set for the speaker-identity conversion task 
consists of utterances of the same set of sentences read by different speakers.
Therefore, the quality of each converted utterance can be objectively evaluated by treating the corresponding target utterance as the ground truth.
To evaluate the similarity between the converted and target utterances, we used the mel-cepstral distortion (MCD) and log $F_0$ correlation coefficient (LFC).

To evaluate the speed of conversion by the BAT version of each method 
and the feasible latency of the RT version of FastS2S-VC, we evaluated the real-time factor (RTF)
and average execution time taken within each sliding window.

\subsubsection{Mel-cepstral distortion}

Given the pair of a converted speech signal and the corresponding reference speech signal, we used the average of the MCDs taken along the dynamic time warping path between the mel-cepstrum sequences of the two signals.
The smaller the MCD, the better the performance.

\subsubsection{Log $F_0$ Correlation Coefficient}
To evaluate the log $F_0$ contour of converted speech, 
we used the LFC \cite{Hermes1998short} between 
the converted and target speech as the objective performance measure. 
In the experiment, 
we used the average of LFCs taken over all the test utterances.
The closer the LFC is to 1, the better the performance.

\subsubsection{Real-Time Factor}

To evaluate the speed of conversion by the BAT version of each method,
we measured RTF, i.e.,
the execution time divided by the length of the input speech.

\subsubsection{Execution Time}

To determine the feasible latency for the RT version of FastS2S-VC, we measured the absolute execution time in milliseconds within each sliding window under different window lengths, when run on the GPU and CPU, respectively. 

\begin{table}[t!]
\centering
\caption{Real-time factors for (1) feature extraction, (2) feature mapping, and (3) waveform generation in S2S-VC (BAT) and FastS2S-VC (BAT).}
\begin{tabular}{r V{3} c|c|c|c|c}
\thline
\multirow{2}{*}{stage}&\multicolumn{3}{c|}{S2S-VC (BAT)}&\multicolumn{2}{c}{FastS2S-VC (BAT)}\\
\cline{2-6}
&RNN    &Conv   &Trans & Conv & Trans \\\thline
(1)  &0.0039 &0.0037 &0.0039 &0.0038 &0.0039 \\
(2)  &0.1173 &0.3575 &0.7585 &0.0048 &0.0072 \\
(3)  &0.0064 &0.0066 &0.0066 &0.0064 &0.0065 \\\hline
total&0.1276 &0.3678 &0.7689 &0.0150 &0.0177 \\\thline
\end{tabular}
\label{tab:rtf_comp}
\end{table}

\begin{table}[t!]
\centering
\caption{Average execution time in milliseconds on GPU and CPU for (1) feature extraction, (2) feature mapping, and (3) waveform generation performed at each sliding window with the RT version of FastS2S-VC. }
\begin{tabular}{r|r V{3} c|c|c|c}
\thline
\multirow{2}{*}{$S$ [ms]}&\multirow{2}{*}{stage}&
\multicolumn{2}{c|}{GPU}&\multicolumn{2}{c}{CPU}
\\\cline{3-6}
       &                &Conv&Trans&Conv&Trans\\\thline
 \multirow{4}{*}{32}&(1)&$2.5$&$2.4$&$2.4$&$2.4$\\
       &             (2)&$9.7$&$16.1$&$88.6$&$85.5$\\
       &             (3)&$13.4$&$13.7$&$56.6$&$56.3$\\\cline{2-6}
       &           total&$\bf 25.5$&$32.2$&$147.6$&$144.2$\\\hline
 \multirow{4}{*}{64}&(1)&$2.5$&$2.5$&$2.5$&$2.6$\\
       &             (2)&$10.1$&$16.2$&$94.8$&$90.6$\\
       &             (3)&$13.9$&$14.5$&$68.9$&$67.7$\\\cline{2-6}
       &           total&$\bf26.6$&$\bf33.1$&$166.2$&$160.9$\\\hline
\multirow{4}{*}{128}&(1)&$2.7$&$2.6$&$2.7$&$2.7$\\
       &             (2)&$10.8$&$16.6$&$95.3$&$92.4$\\
       &             (3)&$14.9$&$16.1$&$91.8$&$90.7$\\\cline{2-6}
       &           total&$\bf 28.4$&$\bf 35.4$&$189.7$&$185.8$\\\hline
\multirow{4}{*}{256}&(1)&$3.3$&$3.3$&$3.4$&$3.4$\\
       &             (2)&$11.1$&$17.2$&$97.6$&$93.6$\\
       &             (3)&$17.0$&$17.6$&$132.1$&$130.1$\\\cline{2-6}
       &           total&$\bf 31.4$&$\bf 38.1$&$\bf 233.1$&$\bf 227.1$\\\thline  
\end{tabular}
\label{tab:et_comp}
\end{table}

\subsection{Objective Evaluations}

\reffig{arctic_b0411} shows 
examples of attention matrices predicted by 
S2S-VC--BAT--Conv (namely, ConvS2S-VC2), the teacher and student models in FastS2S-VC--BAT--Conv,
S2S-VC--BAT--Trans (namely, Transformer-VC2), and the teacher and student models in FastS2S-VC--BAT--Trans. 
From these examples, we can see that the teacher and student models in FastS2S-VC have successfully been able to produce attention matrices similar to those produced by the original S2S-VC in both architectures.

The MCDs and LFCs obtained with all the methods in the speaker-identiy conversion task are shown in \reftab{mcd_baseline_comp}. 
For results obtained by other methods under the same conditions, please refer to \cite{Kameoka2018IEEE-TASLP_ConvS2S-VC,Kameoka2020IEEE-TASLP_VTN,Kameoka2020IEEETrans_AStarGAN-VC}.
Comparing between architectures, the result showed that the convolutional architecture performed best for all the methods, followed by the Transformer architecture.
This result is actually consistent with the result reported in \cite{Kameoka2020IEEE-TASLP_VTN}, indicating that the size of the present training set 
may not have fully exploited the potential
of the Transformer architecture,
and that the convolutional architecture can be a reasonable choice when the amount of training data is limited, as in the present dataset.
Under the same architectural conditions, the baseline S2S-VC performed the best.
This result is reasonable, since the baseline S2S-VC can take advantage of the ability of the AR structure (but with the disadvantage of slow conversion speed).
What is noteworthy here is that both the BAT and RT versions of FastS2S-VC showed comparable or only slightly worse performance than the baseline S2S-VC, despite the NAR structure. 
It is also important to note that the BAT and RT versions of FastS2S-VC showed comparable performance, which means that the sliding-window type conversion process had little negative impact on the overall conversion quality.

The RTF comparison of S2S-VC and FastS2S-VC is shown in \reftab{rtf_comp}. 
As the result shows, FastS2S-VC was able to perform conversion significantly faster than S2S-VC (more than 70 times faster for the convolutional architecture and 100 times faster for the Transformer architecture) thanks to its NAR structure. 
This explains why we named the proposed method FastS2S-VC.
It is worth noting that for both architectures, the feature mapping process took only as much time as the feature extraction and waveform generation processes.

The average execution time within each sliding window taken by the RT version of FastS2S-VC is shown in \reftab{et_comp}.
The execution time that is shorter than the window length $S$ is shown in bold.
When run on the GPU, both architectures were able to complete the feature extraction, feature mapping, and waveform generation processes within most of the tested window lengths (i.e., 32, 64, 128, and 256 ms).
For example, the shortest feasible latency for the convolutional architecture was 32 ms.
However, when run on the CPU, both architectures could not complete all the processes in the time corresponding to some window lengths.
Specifically, both architectures took longer than $S$ to complete all the processes when $S$ was 32, 64, or 128 ms, and the shortest feasible latency was 256 ms. 
However, as can be seen from the breakdown of the execution time listed in \reftab{et_comp}, the waveform generation process accounts for much of the processing time. This suggests that the latency of the system can be further reduced by using a faster waveform generation method.

\begin{figure*}[t!]
\centering
\begin{minipage}[t]{.54\linewidth}
%\centering
\includegraphics[height=5.5cm]{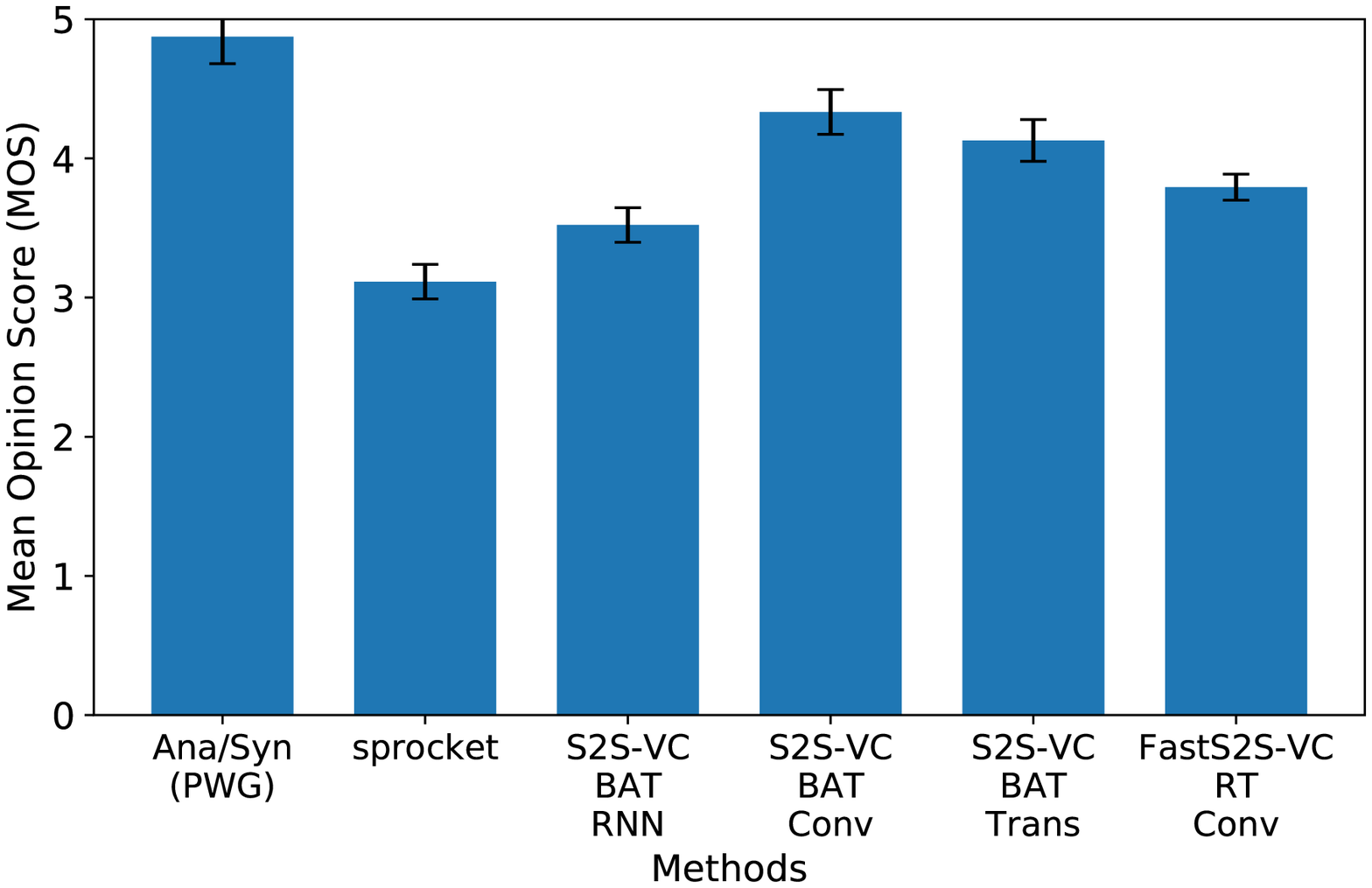}
\vspace{-2ex}
\caption{Audio quality score in speaker-identity conversion task}
\label{fig:audio_quality}
\end{minipage}
\begin{minipage}[t]{.44\linewidth}
\includegraphics[height=5.5cm]{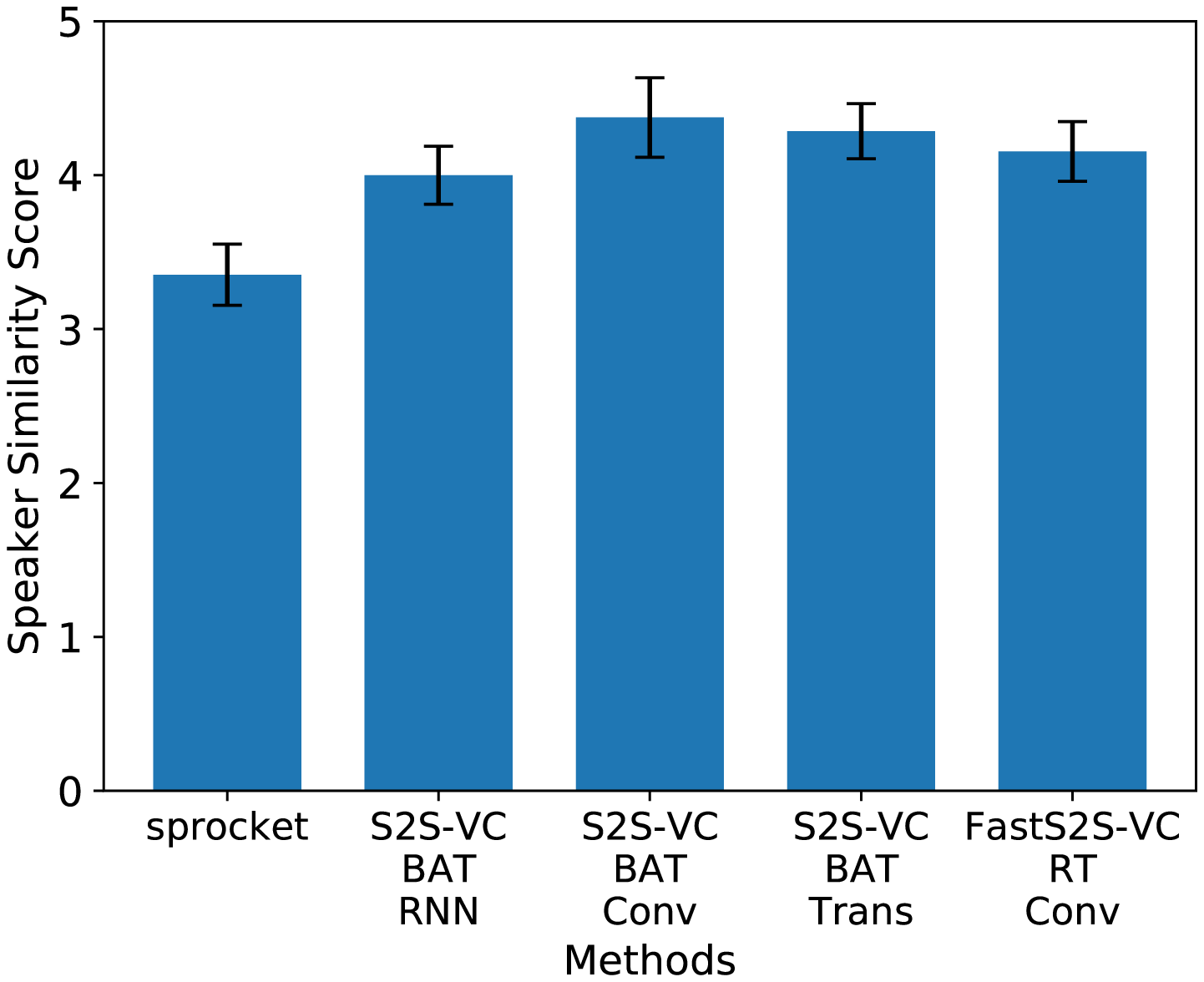}
\vspace{-2ex}
\caption{Similarity score in speaker-identity conversion task}
\label{fig:speaker_similarity}
\end{minipage}
\end{figure*}

\begin{figure}[t!]
\centering
\includegraphics[height=5.5cm]{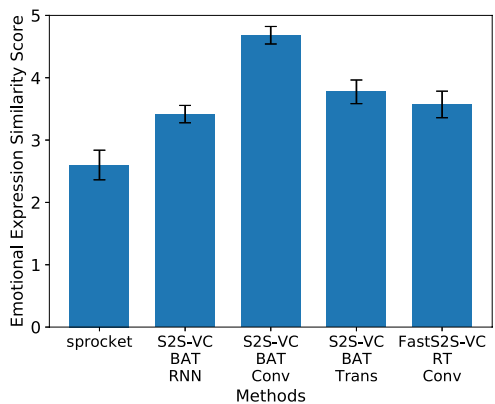}
\vspace{-2ex}
\caption{Similarity score in emotional-expression conversion task}
\label{fig:emotion_similarity}
\end{figure}

\subsection{Subjective Listening Tests}

We conducted subjective listening tests to evaluate 
the audio quality and speaker similarity in the speaker-identity conversion task 
and the emotional expression similarity in the emotional-expression conversion task.
In each test, twenty-four listeners (including 21 native Japanese speakers) participated. 
All the tests were conducted online using Amazon Web Services, and each participant was asked to use a headphone in a quiet environment.
Since the objective evaluation showed that the convolutional version of FastS2S-VC performed better than the Transformer version, and since we are particularly interested in the performance of the real-time version, we used only the audio samples of FastS2S-VC--RT--Conv as representative of FastS2S-VC.

With the audio quality test, the mean opinion score (MOS) was evaluated for each speech sample. 
In this test, we included the audio samples generated by Parallel WaveGAN from the mel-spectrograms extracted from all the test samples. We refer to this method of sample generation as ``Ana/Syn (PWG)''.
Since all the methods compared used Parallel WaveGAN for waveform generation, the scores for these samples can be seen as the upper bound of the performance.
The speech samples were presented in a random order to avoid bias in the order of the stimuli.
Each listener was asked to rate the naturalness of each utterance by selecting 5: Excellent, 4: Good, 3: Fair, 2: Poor, or 1: Bad. 
The obtained scores along with 95\% confidence intervals are shown in \reffig{audio_quality}. 
As the results show, the one that produced the best-sounding speech was S2S-VC--BAT--Conv, followed by S2S-VC--BAT--Trans.
FastS2S-VC--RT--Conv was the next best to these two methods. 
It is worth noting that it performed better than S2S-VC--BAT--RNN, which represents existing S2S model-based VC methods.

For the speaker similarity and emotional-expression similarity tests, each listener was asked to rate the subjective scores on a five-point scale, similar to the audio quality test.
In the speaker similarity test, 
each listener was given a converted speech sample and 
a real speech sample of the corresponding target speaker, 
and asked to rate how likely they were to have been uttered by the same speaker on a scale of 
5: Definitely, 4: Likely, 3: Maybe, 2: Not very likely, and 1: Unlikely.
In the emotional-expression similarity test, 
each listener was given a converted speech sample and 
a real speech sample spoken with the corresponding target emotional expression, 
and asked to rate the similarity of their emotional expressions on a scale of 
5: Very similar, 4: Similar, 3: Fair, 2: Not very similar, and 1: Not alike.
The speaker and emotional-expression similarity scores obtained with these tests along with 95\% confidence intervals are shown in \reffigs{speaker_similarity}{emotion_similarity}. 
The results showed that the order of superiority of the tested methods was the same as in the audio quality test, but the performance difference of each method was smaller in the speaker similarity task and larger in the emotional-expression conversion task.
In terms of both similarity scores, the proposed FastS2S-VC performed better than or comparably to S2S-VC--BAT--RNN. This is very promising, considering the advantage of FastS2S-VC being able to operate as a real-time system.

Audio examples of the tested methods can be found here\footnote{http://www.kecl.ntt.co.jp/people/kameoka.hirokazu/Demos/convs2s-vc2/\\
http://www.kecl.ntt.co.jp/people/kameoka.hirokazu/Demos/transformer-vc2/\\
http://www.kecl.ntt.co.jp/people/kameoka.hirokazu/Demos/fasts2s-vc/}.

\section{Conclusions}

In this paper, we proposed FastS2S-VC, an NAR extesnion of S2S-VC tailored to real-time VC.
Our FastS2S-VC is based on a model that converts one mel-spectrogram into another, and consists of prenet, encoder, attention predictor, predecoder, postdecoder, and postnet.
The attention predictor is designed as a fully convolutional network that produces as intermediate output an attention weight matrix represented using constrained Gaussian functions, and does not require the mel-spectrogram of target speech as input, which is the key to free the entire model from an AR structure and allow for parallelization.
The model is trained to learn to behave similarly to the S2S-VC model based on a teacher-student learning framework. 
We also discussed the idea of implementing FastS2S-VC as a real-time system based on a sliding-window approach. 

The speaker-identity and emotional-expression conversion experiments showed that FastS2S-VC was able to 
speed up the conversion process by 70 to 100 times, while performing comparably or only slightly worse compared to the original S2S-VC.
We also showed that the real-time version of FastS2S-VC can be run with a latency of 32 ms when run on the GPU.
When run on the CPU, however, the feasible latency was 256 ms. 
Even though Parallel WaveGAN is known to be a relatively fast method, the waveform generation process took up much of the processing time within each sliding window.  
In the future, we would like to pursue faster waveform generation methods as well as more lightweight network architectures for the feature mapping model to further reduce the feasible latency.

\section*{Acknowledgments}

This work was supported by JST CREST Grant Number JPMJCR19A3, Japan.

\ifCLASSOPTIONcaptionsoff
  \newpage
\fi

\bibliographystyle{IEEEtran}
\bibliography{Kameoka2021arXiv_FastS2S-VC}

\end{document}